\documentclass[11pt,a4paper]{article}
\pdfoutput=1
\usepackage{amsmath,amssymb,tikz}
\usepackage{graphicx}
\usepackage{cleveref}
\usepackage{wrapfig}
\usepackage{cite,color}
 \allowdisplaybreaks
\def\be{\begin{equation}}
\def\ee{\end{equation}}
\def\ba#1\ea{\begin{align}#1\end{align}}
\def\bg#1\eg{\begin{gather}#1\end{gather}}
\def\bm#1\em{\begin{multline}#1\end{multline}}
\def\bmd#1\emd{\begin{multlined}#1\end{multlined}}

\def\({\left(}
\def\){\right)}
\def\[{\left[}
\def\]{\right]}

\def \be {\begin{equation}}
\def \ee {\end{equation}}
\def \ba {\begin{array}}
\def \ea {\end{array}}
\def \bea{\begin{eqnarray}}
\def \eea{\end{eqnarray}}

\def\bea{\begin{eqnarray}}
\def\eea{\end{eqnarray}}

\newcommand{\bit}{\begin{itemize}}  \newcommand{\eit}{\end{itemize}}
\newcommand{\ben}{\begin{enumerate}}  \newcommand{\een}{\end{enumerate}}

\long\def\symbolfootnote[#1]#2{\begingroup%
\def\thefootnote{\fnsymbol{footnote}}\footnote[#1]{#2}\endgroup}
\newcommand{\sysu}{{\it School of Physics and Astronomy, Sun Yat-Sen University, 2 Daxue Road, Zhuhai 519082, China}}
\begin{document}
%%%%%%%%%%%%%%%%%%%%%%%%%%%%%%%%%%%%%%%%%%%
\thispagestyle{empty}
%%%%%%%%%%%%%%%%%%%%%%%%%%%%%%%%%%%%%%%%%%%
%%%%%%%%%%%%%%%%%%%%%%%%%%%%%%%%%%%%%%%%%%%
%\begin{flushright}
%\hfill{AEI-2015-xxx}
%\end{flushright}
%%%%%%%%%%%%%%%%%%%%%%%%%%%%%%%%%%%%%%%%%%%
\begin{center}

~\vspace{20pt}

{\Large\bf Note on Anomalous Currents for a Free Theory}

\vspace{25pt}

Peng-Ju Hu, Qi-Lin Hu, Rong-Xin Miao\symbolfootnote[2]{Email:~\sf
  miaorx@mail.sysu.edu.cn}

\vspace{10pt}\sysu

\vspace{2cm}

\begin{abstract}
Recently it is found that, due to Weyl anomaly, an external magnetic field can induce anomalous currents near a boundary.  
In this note, we study anomalous currents for complex scalars and Dirac fields in general dimensions.  We develop a perturbation method to calculate Green's function in the spacetime with boundaries. By applying this method, we obtain anomalous currents up to the linear order of magnetic fields in a half space and in a strip.  To the best of our knowledge, the results for Dirac fermions and for strips are new. It is remarkable that, unlike the scalars and holographic BCFT, the anomalous currents of Dirac fields are independent of boundary conditions in general dimensions. Besides, the currents of Dirac fields are always larger than those of complex scalars. Finally, we find an exact formal expression of the anomalous current in a half space. The result is expressed in momentum integrals, which can be evaluated numerically. We find that the mass suppresses the anomalous currents as expected. 
\end{abstract}

\end{center}

%%%%%%%%%%%%%%%%%%%%%%%%%%%%%%%
\newpage
\setcounter{footnote}{0}
\setcounter{page}{1}
%%%%%%%%%%%%%%%%%%%%%%%%%%%%%%%

\tableofcontents
%%%%%%%%%%%%%%%%%%%%%%%%%%%%%%%

\section{Introduction}

The anomaly-induced transport is an important phenomenon, which has a wide range of applications \cite{review}.  The well-known examples include chiral magnetic effect (CME) \cite{Vilenkin:1995um,
Vilenkin:1980fu, Giovannini:1997eg, alekseev, Fukushima:2012vr} and chiral vortical effect (CVE) \cite{Kharzeev:2007tn,Erdmenger:2008rm,
 Banerjee:2008th,Son:2009tf,Landsteiner:2011cp,Golkar:2012kb,Jensen:2012kj}, which refer
to the generation of currents due to an external magnetic field and the rotational motion in the charged fluid, respectively.  It is interesting that the CME current is topologically protected and hence non-dissipative \cite{review}. Similar to chiral anomaly, Weyl anomaly \cite{Duff:1993wm} can also induce anomalous currents in an external electromagnetic field \cite{Chernodub:2016lbo, Chernodub:2017jcp,Chu:2018ksb,Chu:2018ntx}. See \cite{Chu:2018fpx,Chu:2019rod,Miao:2017aba,Miao:2018dvm,Chernodub:2018ihb,Chernodub:2019blw,Ambrus:2019khr,Zheng:2019xeu} for related works. It is remarkable that a similar mechanism leads to novel Fermi condensations when a background scalar field is turned on  \cite{Chu:2020mwx}. The scalar field can either be the Higgs field in a fundamental theory or the phonon in condensed matter system.

On the other hand, the boundary effect of quantum field theory is another interesting phenomenon. Famous examples include Casimir effect \cite{Casimir:1948dh,Plunien:1986ca,Bordag:2001qi} and topological insulator \cite{Hasan:2010xy}. Recently, much attention has been drawn on boundary conformal field theory (BCFT) \cite{Cardy:2004hm,McAvity:1993ue} and its holographic dual (AdS/BCFT) \cite{Takayanagi:2011zk} . Please see \cite{Chu:2018fpx,Chu:2019rod,Miao:2017aba,Miao:2018dvm,Jensen:2015swa,Fursaev:2015wpa,Herzog:2015ioa,Herzog:2017kkj,Herzog:2017xha,Jensen:2017eof,Kurkov:2017cdz,Kurkov:2018pjw,Vassilevich:2018aqu,Fialkovsky:2019rum,Vassilevich:2019mhl,Rodriguez-Gomez:2017kxf,Berthiere:2019lks,FarajiAstaneh:2017hqv,Seminara:2017hhh,Erdmenger:2015spo,Erdmenger:2014xya,Flory:2017ftd,Miao:2017gyt,Chu:2017aab,Miao:2018qkc,Herzog:2019bom,Herzog:2019rke} for some of recent developments. It is interesting that a-Type anomalies of BCFT/ dCFT can depend on marginal couplings \cite{Herzog:2019rke}.

% v2
Weyl anomaly measures the breaking of scaling symmetry of CFT/BCFT due to quantum effects \cite{Duff:1993wm}. It is closely related to the UV Logarithmic divergent term of effective action \cite{Chu:2018ksb,Miao:2017aba}. As a result, one can derive a key relation  \cite{Chu:2018ksb}
\begin{eqnarray}\label{keyrelation}
(\delta \mathcal{A})_{\partial M}=\left( \int_M \sqrt{|g|} <J^{\mu}> \delta A_{\mu} \right)_{\log \epsilon},
\end{eqnarray}
between the renormalized current $<J^{\mu}>$ and the boundary part of the variation of the Weyl anomaly $\mathcal{A}$. Here  $A_{\mu}$ is the gauge field and $\epsilon$ denotes the cutoff of the theory. 
By applying (\ref{keyrelation}), \cite{Chu:2018ksb,Chu:2018ntx} find that, due to the Weyl anomaly, 
 \begin{eqnarray}\label{anomaly}
\mathcal{A}= \int_M \sqrt{|g|} \ \frac{\beta}{2} F^{\mu\nu}F_{\mu\nu},
\end{eqnarray}
 an external electromagnetic field can induce an universal current
\begin{eqnarray}\label{typeIIcurrent}
<J^{\mu}>=\frac{-2\beta F^{\mu\nu}n_{\nu}}{x} +..., \ x\sim 0,
\end{eqnarray}
near the boundary, where $\beta$ is the beta function, $F^{\mu\nu}$ are the field strength, $x$ is the proper distance to the boundary, $n_i$ are the normal vectors and $...$ denote higher order terms in $O(x)$.  Note that there are boundary contributions to the current, which can cancel the `bulk' divergence of 
(\ref{typeIIcurrent}) and make finite the total current \cite{Chu:2018ksb} .  It is remarkable that the leading term of anomalous current (\ref{typeIIcurrent}) is universal in four dimensions. It is independent of the boundary conditions (BC), the temperature and the details of theories. What is more, it applies to not only conformal field theory but also the general quantum field theory \cite{Chu:2018ksb}. In higher dimensions, the anomalous current is expected to take the following form  \cite{Chu:2018ksb}
\begin{eqnarray}\label{typeIIcurrentd}
<J_d^{\mu}>=\frac{b_d \ F^{\mu\nu}n_{\nu}}{x^{d-3}} +..., \ x\sim 0,
\end{eqnarray}
where $b_d$ are central charges of Weyl anomaly and $d$ denote the dimensions. Note that $b_d$ depend on boundary conditions in dimensions other than four. 

In this paper, we investigate the anomalous current in a spacetime with a boundary. 
The previous works \cite{Chu:2018ksb,Chu:2018ntx} mainly discuss the general characteristics of anomalous currents.  In particular, they focus on the region close to the boundary. In this note, we study more specific examples such as complex scalars and Dirac fermions, and try to explore the anomalous current in the full region of the system.  By applying Green's function method, we derive the anomalous currents for free theories up to the linear order of external magnetic field in a half space and in a strip. Our results agree with the work of \cite{McAvity:1990we} for complex scalars in a half space. To the best of our knowledge, the anomalous currents for Dirac fermions and for strips are new. Our results are exact in the the size of strip, hence works well in the full region of a strip.  We also obtain an exact formal expression of the anomalous current for complex scalars in a half space. The formal formula is expressed in momentum integrals,  which can be evaluated numerically. Let us summarize the properties of anomalous currents for free theories below.
 
{\tt 1}.  Unlike the holographic BCFT \cite{Chu:2018ntx} and complex scalars, the anomalous currents of Dirac fields are independent of BCs in general dimensions. 

{\tt 2}.  The anomalous currents of complex scalars have different signs for Dirichlet boundary condition (DBC) and Neumann boundary condition (NBC) in dimensions higher than four. See Figure. 2 for example. 

{\tt 3}.  The anomalous currents of Dirac fields are larger than those of complex scalars in four dimensions. See Figure. 3 for example. 

%{\tt 4}.  In a strip, the anomalous currents become larger as the dimensions increase. See Figure. 4 for example.

{\tt 4}.  The mass suppresses the anomalous current and the current approaches zero far aways from the boundary.  See Figure. 5 for example.

The paper is organized as follows. 
In sect. 2, we develop a perturbation method to calculate Green's function. Our method includes only bulk integral, which is slightly different from that of \cite{Deutsch:1978sc,JohnP}. In sect. 3, by applying the perturbation method, we derive the anomalous currents for complex scalar and Dirac field in a half space in general dimensions. In sect. 4,  we study the anomalous current in a strip. In sect. 5,  we obtain a formal expression of the anomalous current in a half space. Finally, we conclude with discussions in sect. 6.  We use conventions of \cite{Parker:2009uva} and the signature of metric is $(1,-1,...,-1)$.

\section{Green's Function}

Green's function is a powerful tool to calculate expectation value of stress-energy tensors and currents near a boundary  \cite{Deutsch:1978sc,JohnP}. Usually, Green's function is expressed as a boundary integral for BCFT \cite{Deutsch:1978sc,JohnP}. Here we develop a slightly different approach where Green's function includes only bulk integrals. Our approach has the advantage that the nth term of series of Green's function is of order $O(B^n)$
\begin{eqnarray}\label{Series}
G=\sum_{n=0}^{\infty} G_n, \ \ \ \ G_n \sim O(B^n)
\end{eqnarray}
where $B$ denote the magnetic field or other perturbation parameters. As a result, to derive the leading term of anomalous currents of $O(B)$, we only need to calculate one term $G_1$ in the Green's function.  To illustrate our approach, let us take complex scalars and Dirac fields as examples below.

\subsection{Complex scalar }
Let us start with the action of free complex scalars in a curved spacetime
\begin{eqnarray}\label{action}
I=\int_M dx^d \sqrt{|g|}D_{\nu} \phi (D^{\nu} \phi)^* ,
\end{eqnarray}
where $d$ is the dimension of spacetime and $D_{\mu}=\nabla_{\mu}- i e A_{\mu}$ is the covariant derivative. For simplicity, we set $e=1$ in this paper. The Green function satisfies equation of motion (EOM)
\begin{eqnarray}\label{GreenEOM}
D^{\mu}D_{\mu} G(x,x')=(\nabla^{\mu}\nabla_{\mu} + E)G(x,x')=\delta(x,x'),
\end{eqnarray}
where we take $E=(-2i  A^{\mu}\nabla_{\mu}-i  \nabla^{\mu} A_{\mu}-A^{\mu}A_{\mu})$ as perturbations.   One can impose either DBC
\begin{eqnarray}\label{GreenDBC}
G(x,x')|_{\partial M}=0,
\end{eqnarray}
or NBC
\begin{eqnarray}
D_n G(x,x')|_{\partial M}=0, \label{GreenNBC}
\end{eqnarray}
on the boundary  $\partial M$. Here $n$ denotes the normal direction.  

We split Green's function into the background $G_0$ and a correction $G_c$
\begin{eqnarray}  \label{Greentotal}
G=G_0+G_c,
\end{eqnarray}
where $G_0$ obeys EOM
\begin{eqnarray}\label{G0EOM}
\nabla^{\mu}\nabla_{\mu} G_0(x,x')=\delta(x,x'),
\end{eqnarray}
together with either DBC
\begin{eqnarray}\label{G0DBC}
G_0|_{\partial M}=0,
\end{eqnarray}
or NBC
\begin{eqnarray}
\nabla_n G_0|_{\partial M}=0. \label{G0NBC} 
\end{eqnarray}
Applying Green's formula, we have
\begin{eqnarray}  \label{Greenformula}
&&\int_M dx^d\sqrt{|g|} \Big[  G_c(x', x) D^{\mu} D_{\mu} G(x, x'') -G_c(x', x)  \overleftarrow{D}^{*\mu}  \overleftarrow{D}^*_{\mu} G(x, x'') \Big]\nonumber\\
&=&\int_{\partial M} dy^{d-1} \sqrt{|h|} \Big[  G_c(x', x) D_{n} G(x, x'')- G_c(x', x) \overleftarrow{D}^*_{n} G(x, x'') \Big]
\end{eqnarray}
where $ \overleftarrow{D}^*_{\mu} =\overleftarrow{\nabla}_{\mu} +i  A_{\mu}$ means acting on the left. We choose the gauge $A_n=0$ so that $D_{n}=D^*_{n} =\nabla_n$ on the boundary. 
  Imposing either DBC (\ref{GreenDBC}, \ref{G0DBC}) or NBC (\ref{GreenNBC}, \ref{G0NBC}), we find that the boundary terms of (\ref{Greenformula}) vanish. From EOM (\ref{GreenEOM}, \ref{G0EOM}), we derive
\begin{eqnarray}\label{GcEOM}
G_c(x', x)  \overleftarrow{D}^{*\mu}  \overleftarrow{D}^*_{\mu} = -G_0 (x', x) \overleftarrow{E}^{*} ,
\end{eqnarray}
where 
\begin{eqnarray}\label{E}
\overleftarrow{E}^*=(2i \overleftarrow{\nabla}_{\mu}A^{\mu}+i  \nabla^{\mu} A_{\mu}- A^{\mu}A_{\mu}).
\end{eqnarray}
 Substituting (\ref{GreenEOM}) and (\ref{GcEOM}) into (\ref{Greenformula}) and noting that the boundary terms vanishing due to BCs,  we obtain a key formula
\begin{eqnarray}\label{GcSolution}
G_c(x', x'')=-\int_M dx^d \sqrt{|g|} \Big[ G_0(x', x) \overleftarrow{E}^*(x) G(x,x'')  \Big].
\end{eqnarray} 
Unlike \cite{Deutsch:1978sc,JohnP}, there are only bulk integrals in $G_c$. From (\ref{GcSolution}), we can calculate $G_c$ perturbatively
\begin{eqnarray}\label{Gcper}
G_c(x', x'')&=&-\int_M dx^d\sqrt{|g|} G_0(x', x) \overleftarrow{E}^*(x) G_0(x,x'')  \nonumber\\
&&+ \int_M dx^d\sqrt{|g|}  \int_{M_1} dx_1^d\sqrt{|g_1|} G_0(x', x) \overleftarrow{E}^*(x) G_0(x,x_1)  \overleftarrow{E}^*(x_1) G_0(x_1,x'') \nonumber\\
&& +...
\end{eqnarray} 
where the nth line of (\ref{Gcper}) is of oder $O(E^n)$. 

\subsection{Dirac field}

Now let us turn to study Green's function of Dirac fields.  The action of free Dirac fields in a curved spacetime is given by
\begin{eqnarray}\label{action}
I=\int_M dx^d\sqrt{|g|}\bar{\Psi} i \gamma^{\mu} D_{\mu} \Psi,
\end{eqnarray}
where $\bar{\Psi}=\Psi^+\gamma^0$ and $D_{\mu}=\nabla_{\mu}-i  A_{\mu}$. 
Green's function obeys EOM 
\begin{eqnarray}\label{GreenEOMDirac}
i \gamma^{\mu} D_{\mu} S(x,x')=\delta(x,x'),
\end{eqnarray}
where $\delta(x,x')=\delta^d(x-x')/\sqrt{|g|}$.   We impose bag boundary condition (BBC)
\begin{eqnarray}\label{GreenBBC}
\Pi_{-} S(x,x')|_{\partial M}=0,
\end{eqnarray}
where $\Pi_{\pm}=(1\pm \chi)/2$ are projection operators and $\chi$ satisfy \cite{McAvity:1993ue}
\begin{eqnarray}\label{gamma}
\chi \gamma_n=-\gamma_n \bar{\chi}, \  \chi \gamma_a=\gamma_a \bar{\chi}, \ \chi^2=\bar{\chi}^2=1,
\end{eqnarray}
where $\bar{\chi}=\gamma^0 \chi^+ \gamma^0$ and $n$ ($a$) denote the normal (tangent) directions.  From BBC (\ref{GreenBBC},\ref{gamma}), we can derive
\begin{eqnarray}\label{GreenBBC1}
S(x'',x) \gamma_n S(x,x')|_{\partial M}=0.
\end{eqnarray}

We split Green function into a background $S_0$ and a correction $S_c$
\begin{eqnarray}  \label{GreentotalDirac}
S=S_0+S_c,
\end{eqnarray}
where $S_0$ obeys EOM
\begin{eqnarray}\label{S0EOM}
i \gamma^{\mu} \nabla_{\mu}  S_0(x,x')=\delta(x,x'),
\end{eqnarray}
together with BBC
\begin{eqnarray}\label{S0BBC}
\Pi_{-} S_0(x,x')|_{\partial M}=0.
\end{eqnarray}
From (\ref{GreenBBC},\ref{gamma},\ref{S0BBC}), we have
\begin{eqnarray}\label{BBCBBC}
S_A(x'',x) \gamma_n S_B(x,x')|_{\partial M}=0.
\end{eqnarray}
where $S_{A,B}$ denote $S,S_0, S_c$. 

Applying Green's formula for Dirac fields, we have
\begin{eqnarray}  \label{GreenformulaDirac}
&&\int_M dx^d\sqrt{|g|} \Big[  S_c(x', x) i \gamma^{\mu}( \overrightarrow{\nabla}_{\mu}-i  A_{\mu} ) S(x, x'') + S_c(x', x) i \gamma^{\mu}( \overleftarrow{\nabla}_{\mu}+ i  A_{\mu} ) S(x, x'') \Big]\nonumber\\
&=&\int_{\partial M} dx^{d-1}\sqrt{|h|} \Big[  S_c(x', x) i \gamma_n S(x, x'') \Big].
\end{eqnarray}
Note that the boundary term of (\ref{GreenformulaDirac}) vanishes due to (\ref{BBCBBC}).  Applying EOM (\ref{GreenEOMDirac}, \ref{S0EOM}), we derive
\begin{eqnarray}\label{ScEOM}
S_c(x', x) i \gamma^{\mu}( \overleftarrow{\nabla}_{\mu}+ i A_{\mu} )=S_0(x',x) \gamma^{\mu}A_{\mu}
\end{eqnarray}
Now (\ref{GreenformulaDirac})  can be simplified as
\begin{eqnarray}\label{ScSolution}
S_c(x', x'')=-\int_M dx^d\sqrt{|g|} \Big[ S_0(x', x) \gamma^{\mu}(x)A_{\mu}(x) S(x,x'')  \Big].
\end{eqnarray} 
 From (\ref{ScSolution}), we can calculate $S_c$ perturbatively
\begin{eqnarray}\label{Scper}
S_c(x', x'')&=&-\int_M dx^d\sqrt{|g|} S_0(x', x) \gamma^{\mu}A_{\mu}(x) S_0(x,x'')  \nonumber\\
&+& \int_M dx^d \sqrt{|g|} \int_{M_1} dx_1^d \sqrt{|g_1|} S_0(x', x)  \gamma^{\mu}(x)A_{\mu}(x) S_0(x,x_1)\gamma^{\mu}(x_1)A_{\mu}(x_1) S_0(x_1,x'') \nonumber\\
&+&...
\end{eqnarray} 
where the nth line of (\ref{Scper}) is of oder $O(A^n)$.

Now we finish the  perturbative derivations of Green's function for complex scalars (\ref{Gcper}) and Dirac fields (\ref{Scper}).

\section{Current in a half space}

In this section we calculate the anomalous current for complex scalars and Dirac fields in a half space. For simplicity, we focus on the half space $x \ge 0$ with a constant magnetic field parallel to the boundary. We have $x^{\mu}=(t, x, y_a)=(t, x, y_1, ..., y_{d-2})$,  $A_{\mu}=(0,0, B x, 0,..., 0)$ and $g_{\mu\nu}=\eta_{\mu\nu}=\text{diag}(1,-1,....,-1)$. 

\subsection{Complex scalar}

Green's function of the complex scalar is given by \cite{Deutsch:1978sc}
\begin{eqnarray}\label{Gscalar0}
G(x,x')=i  <T \phi(x)\phi^*(x')>
\end{eqnarray} 
where $T$ is the time-ordering symbol.  The non-renormalized current is defined by
\begin{eqnarray}\label{nonrenoJ}
\hat{J}_{\mu}(x)=\frac{\delta I_{eff}}{\sqrt{|g|}\delta A^{\mu}}=\lim_{x'\to x}  ( D_{\mu} -D^*_{\mu'} ) G(x,x') 
\end{eqnarray} 
which is divergent generally. Here $ I_{eff}$ denotes the effective action. To get the renormalized current $J_{\mu}$, one should subtract the reference current without boundaries
\begin{eqnarray}\label{renoJ}
J_{\mu}(x)=\lim_{x'\to x}  ( D_{\mu} -D^*_{\mu'} ) \left( G(x,x')-\bar{G}(x,x') \right),
\end{eqnarray} 
where $\bar{G}$ is reference Green's function in the spacetime without boundaries.
By using (\ref{Gcper}), we can obtain $G$ and $\bar{G}$ as
\begin{eqnarray}\label{Gscalar}
G(x', x'')=G_0(x', x'')+2i B \int_0^{\infty}dx \int_{-\infty}^{\infty}dtdy^{d-2} G_0(x', x) x \overleftarrow{\partial}_{y_1}G_0(x,x'')+O(B^2), \\
 \bar{G}(x', x'')=\bar{G}_0(x', x'')+2i  B \int_{-\infty}^{\infty}dx \int_{-\infty}^{\infty}dtdy^{d-2} \bar{G}_0(x', x) x \overleftarrow{\partial}_{y_1}\bar{G}_0(x,x'')+O(B^2). \label{barGscalar}
\end{eqnarray} 
Note that the integral regions of $x$ are different for $G$ and $\bar{G}$.  Here $G_0$ and $\bar{G}_0$ are Green's functions without external magnetic fields
\begin{eqnarray}\label{G0scalar}
&&G_0(x', x'')=\frac{\Gamma \left(\frac{d}{2}-1\right)}{4 \pi ^{d/2}} \Big( \frac{i}{[(x'-x'')^2+(y'_a-y''_a)^2-(t'-t'')^2]^{\frac{d-2}{2}}} \nonumber\\
&&\ \ \ \ \ \ \ \ \ \ \ \ \ \ \ \ \ \ \ \ \ \ \ \ \ \ \ \ \ +\chi \frac{i}{[(x'+x'')^2+(y'_a-y''_a)^2-(t'-t'')^2]^{\frac{d-2}{2}}} \Big), \\
&& \bar{G}_0(x', x'')=\frac{\Gamma \left(\frac{d}{2}-1\right)}{4 \pi ^{d/2}}\frac{i}{[(x'-x'')^2+(y'_a-y''_a)^2-(t'-t'')^2]^{\frac{d-2}{2}}},\label{barG0scalar}
\end{eqnarray} 
where $\chi=-1$ for DBC and $\chi=1$ for NBC. One can check that $G_0$ (\ref{G0scalar}) satisfy EOM (\ref{G0EOM})
and BCs (\ref{G0DBC},\ref{G0NBC}).  

To do the integral along $t$, it is more convenient to perform the Wick rotation $t= -i t_E$.
Substituting (\ref{Gscalar}-\ref{barG0scalar}) into (\ref{renoJ}) and performing the Wick rotation, we get
\begin{eqnarray}\label{scalarintegral1}
J_{y_1}= \chi \ 2^{1-d} \pi ^{-\frac{d}{2}} \Gamma \left(\frac{d}{2}-1\right) B x^{3-d} +B \int_{0}^{\infty} dx' \int_0^{\infty} dr (f_1+ \chi \ f_2)+O(B^2),
\end{eqnarray} 
where
\begin{eqnarray}\label{scalarf1}
f_1=\frac{(2-d)  r^{d-2} \Gamma \left(\frac{d}{2}-1\right)^2 x' \left((d-3) r^2+(d-1) (x+x')^2\right)}{4\pi ^{\frac{1}{2} (d+1)} \Gamma \left(\frac{d+1}{2}\right) \left(r^2+\left(x'+x\right)^2\right)^{d}}
\end{eqnarray} 
\begin{eqnarray}\label{scalarf2}
&&f_2=\frac{x' (2-d) \pi ^{-\frac{d}{2}-\frac{1}{2}} r^{d-2} \Gamma \left(\frac{d}{2}-1\right)^2}{4 \Gamma \left(\frac{d+1}{2}\right) \left(r^2+\left(x-x'\right)^2\right)^{\frac{d}{2}+1} \left(r^2+\left(x'+x\right)^2\right)^{\frac{d}{2}+1}}\Big[ (d-3) r^6+(3 d-7) r^4 \left(x^2+x'^2\right)\nonumber\\
&&+r^2 \left((3 d-5) x^4+(3 d-5) x'^4+2 (1-3 d) x^2 x'^2\right)+(d-1) \left(x^2-x'^2\right)^2 \left(x^2+x'^2\right) \Big]
\end{eqnarray} 
Here $r^2=y_a^2+t_E^2$, $\chi^2=1$ and we have performed angle integrals above. After the radial integration, we get
\begin{eqnarray}\label{scalarintegral2}
 \int_0^{\infty} dr (f_1+ \chi f_2)=\begin{cases}
\frac{ 2^{2-d} \pi ^{-\frac{d}{2}} \Gamma \left(\frac{d}{2}\right) x' \left(\left(x'\right)^{d-1}+\chi  \left(x'+x\right)^{d-1}\right)}{(1-d) \left(x'\right)^{d-1} \left(x'+x\right)^{d-1}}, \ \ x'>x\\
\frac{2^{2-d} \pi ^{-\frac{d}{2}} \Gamma \left(\frac{d}{2}\right) x' \left(x^{d-1}+\chi  \left(x'+x\right)^{d-1}\right)}{(1-d) x^{d-1} \left(x'+x\right)^{d-1}},
 \ \ \ \ \ x'<x.
\end{cases}
\end{eqnarray}
Note that the integrals are different for $x'>x$ and $x'<x$. 
Substituting (\ref{scalarintegral2}) into (\ref{scalarintegral1}) and integrating along $x'$, we finally obtain the anomalous current
\begin{eqnarray}\label{scalarcurrentd}
J_{y_1}= \frac{\pi ^{-\frac{d}{2}}  \Gamma \left(\frac{d}{2}-1\right)}{2^d(d-3) (1-d)}\Big(2-(d-4) (d-1) \chi \Big)\frac{B}{x^{d-3}}+O(B^2),
\end{eqnarray} 
which agree with results of \cite{McAvity:1990we} derived by the heat-kernel method. Recall that $\chi=1$ ($\chi=-1$) for NBC (DBC). The other components of currents vanish. 
 Some comments are in order. First, the anomalous current depends on BCs and have different sign for different BCs in dimensions other than four. Second, it is remarkable that the current is independent of BC in four dimensions
\begin{eqnarray}\label{currentscalar}
J_{4d}^{\mu}=\delta^{\mu}_{ y_1}\frac{B}{24 \pi ^2 x}+O(B^2),
\end{eqnarray} 
which agrees with the result (\ref{typeIIcurrent})  \cite{Chu:2018ksb,Chu:2018ntx}. Note that $J^{y_1}=-J_{y_1}$ in our conventions. Third, in the above calculations, we have assumed $d>3$.  After the analytical extension, the result (\ref{scalarcurrentd}) works well for DBC in two and three dimensions. However, this is not the case for NBC. In dimensions lower than four, we have
\begin{eqnarray}\label{scalarcurrent2d}
J_{2d}^{\mu}=\delta^{\mu}_{y_1}\begin{cases}
\frac{B x}{2 \pi }+O(B^2), \ \ \ \ \ \ \ \ \ \ \ \ \ \ \ \ \ \ \ \ \ \ \ \ \ \ \ \ \ \ \ \ \ \ \ \ \ \ \ \ \ \ \ \ \ \ \ \ \ \ \ \ \ \ \  \text{DBC}\\
\frac{2 B x}{\pi  (2-d)}+\frac{B x (4 \log (x)+2 \gamma -1+\log (16\pi^2))}{2 \pi }+O\left(d-2,B^2\right), \ \ \ \ \ \ \text{NBC},
\end{cases}
\end{eqnarray}
\begin{eqnarray}\label{scalarcurrent3d}
J_{3d}^{\mu}=\delta^{\mu}_{y_1}\begin{cases}
\frac{B}{16 \pi }+O(B^2), \ \ \ \ \ \ \ \ \ \ \ \ \ \ \ \ \ \ \ \ \ \ \ \ \ \ \ \ \ \ \ \ \ \ \ \ \ \ \ \ \ \ \ \ \ \ \ \ \ \ \ \ \ \ \ \ \ \ \text{DBC}\\
\frac{B}{4 \pi  (d-3)}-\frac{B \left(4 \log (x)+3+\log (16 \pi^2)-2 \psi ^{(0)}\left(\frac{1}{2}\right)\right)}{16 \pi }+O\left(d-3,B^2\right), \ \ \text{NBC},
\end{cases}
\end{eqnarray}
where $\gamma$ is Euler Gamma function and $\psi $ is the PolyGamma function. One may regularize the currents for NBC by 
\begin{eqnarray}\label{regularize2d}
&&J^{\mu}_{2d}=\lim_{\epsilon\to 0}\frac{J^{\mu}(d=2+\epsilon)+J^{\mu}(d=2-\epsilon)}{2} \\
&&J^{\mu}_{3d}=\lim_{\epsilon\to 0}\frac{J^{\mu}(d=3+\epsilon)+J^{\mu}(d=3-\epsilon)}{2}. \label{regularize3d}
\end{eqnarray} 
In this way, the divergences of (\ref{scalarcurrent2d},\ref{scalarcurrent3d}) cancel and we get finite currents for NBC in two and three dimensions. 
 We leave a careful discussion of the lower-dimensional currents to future work.

\subsection{Dirac field}

Let us go on to discuss the anomalous current for Dirac field.  Similarly, we have $A_{\mu}=(0,0,B x, 0,...,0)$. Without loss of generality, we choose the chiral bag boundary condition \cite{Goldhaber}
\begin{eqnarray}\label{ChiralBC}
\Pi_{-} \Psi|_{\partial M}=\frac{1+i e^{i \theta \gamma_5 } \gamma^{n}}{2} \Psi|_{\partial M}=0,
\end{eqnarray}
where $\theta$ is a constant and $n$ denotes the normal direction, i.e., $\gamma^n=\gamma^1$.  Equivalently, we have chosen 
\begin{eqnarray}\label{ChiralX}
\chi=-i e^{i \theta \gamma_5 } \gamma^{n}.
\end{eqnarray}
One can check that it satisfies the conditions (\ref{gamma}). Note that (\ref{ChiralBC}) reduces to the usual bag boundary condition  when $\theta=0,\pi$. 
% v2
Note also that since $\gamma_5$ is not well defined in odd  dimensions, we set $\theta=0,\pi$ so that the boundary condition (\ref{ChiralBC}) becomes $(1\pm i \gamma^{n}) \Psi|_{\partial M}=0$ in odd dimensions. 

The Feynman Green function of Dirac field is given by \cite{Parker:2009uva}
\begin{eqnarray}\label{GreenDirac}
S(x,x')=-i <T\Psi(x)\bar{\Psi}(x')>,
\end{eqnarray}
from which one can derive the current 
\begin{eqnarray}\label{regCurrentDirac}
J^{\mu}=-i\lim_{x'\to x} \text{Tr}\Big[\gamma^{\mu}\left(S(x,x')-\bar{S}(x,x')\right)\Big],
\end{eqnarray}
where we have subtracted the reference Green function $\bar{S}$ without boundaries.  From the key formula (\ref{Scper}), we get
\begin{eqnarray}\label{SDirac}
S(x', x'')=S_0(x', x'')- B \int_0^{\infty}dx \int_{-\infty}^{\infty}dtdy^{d-2} S_0(x', x) \gamma^2 x S_0(x,x'')+O(B^2), \\
 \bar{S}(x', x'')=\bar{S}_0(x', x'')- B \int_{-\infty}^{\infty}dx \int_{-\infty}^{\infty}dtdy^{d-2} \bar{S}_0(x', x) \gamma^2 x \bar{S}_0(x,x'')+O(B^2). \label{barSDirac}
\end{eqnarray} 
where $\gamma^2=\gamma^{y_1}$ and
\begin{eqnarray}\label{S0Dirac}
&&S_0(x', x'')=\frac{ (d-2) \Gamma \left(\frac{d-2}{2}\right)}{4\pi ^{\frac{d}{2}}} \Big( \frac{\gamma^0(t'-t'') -\gamma^1 (x'-x'')-\gamma^a (y'_a-y''_a)}{[(x'-x'')^2+(y'_a-y''_a)^2-(t'-t'')^2]^{\frac{d}{2}}}\nonumber\\
&&\ \ \ \ \ \ \ \ \  \ \ \ \ \ \ \ \ \ \ \ \ \ \ \ \ \ \ \ \ \ \ \ \ +\chi. \frac{\gamma^0(t'-t'') -\gamma^1 (-x'-x'')-\gamma^a (y'_a-y''_a)}{[(x'+x'')^2+(y'_a-y''_a)^2-(t'-t'')^2]^{\frac{d}{2}}}\Big), \nonumber \\
&& \\
&& \bar{S}_0(x', x'')=\frac{ (d-2) \Gamma \left(\frac{d-2}{2}\right)}{4\pi ^{\frac{d}{2}}}  \frac{\gamma^0(t'-t'') -\gamma^1 (x'-x'')-\gamma^a (y'_a-y''_a)}{[(x'-x'')^2+(y'_a-y''_a)^2-(t'-t'')^2]^{\frac{d}{2}}}.\label{barS0Dirac}
\end{eqnarray} 
Substituting (\ref{SDirac}-\ref{barS0Dirac}) into (\ref{regCurrentDirac}) and performing Wick rotation, we derive
\begin{eqnarray}\label{Diracintegral1}
J^{y_1}= B \int_{0}^{\infty} dx' \int_0^{\infty} dr \frac{r^{d-2} \Gamma \left(\frac{d}{2}\right)^2 x' 2^{\left[\frac{d}{2}\right]-1} \left((d-3) r^2+(d-1) \left(x'+x\right)^2\right)}{\pi ^{\frac{d+1}{2}} \Gamma \left(\frac{d+1}{2}\right) \left(r^2+\left(x'+x\right)^2\right)^d} +O(B^2),
\end{eqnarray} 
where $[\ ]$ denotes the integer part. After the integrals along $x'$ and $r$, we obtain the anomalous current in a half space for Dirac fields
\begin{eqnarray}\label{Diraccurrentd}
J^{\mu}=\delta^{\mu}_{y_1}\frac{ \pi ^{-\frac{d}{2}}  \Gamma \left(\frac{d}{2}\right) 2^{\left[\frac{d}{2}\right]-d+1}}{(d-3)(d-1)}\frac{B}{x^{d-3}}+O(B^2).
\end{eqnarray} 
To the best of our knowledge, this result is new. 
It is remarkable that the leading term of current (\ref{Diraccurrentd}) is independent of choices of BCs (\ref{ChiralBC}), that different chiral angles $\theta$ yield the same current.  This is quite different from the currents of complex scalars (\ref{scalarcurrentd}) and holographic BCFT \cite{Chu:2018ntx} which depend on BCs. To end this section, let us list the currents for Dirac fermions (\ref{Diraccurrentd}) in four dimensions
\begin{eqnarray}\label{Diraccurrent4d}
J_{4d}^{\mu}=\delta^{\mu}_{y_1} \frac{B}{6 \pi ^2 x} + O(B^2),
\end{eqnarray} 
in two  dimensions
\begin{eqnarray}\label{Diraccurrent2d}
J_{2d}^{\mu}=-\delta^{\mu}_{y_1}\frac{B x}{\pi } + O(B^2),
\end{eqnarray} 
and in three dimensions
\begin{eqnarray} \label{Diraccurrent3d}
J_{3d}^{\mu}=\delta^{\mu}_{y_1}\Big(\frac{B}{8 \pi  (d-3)}-\frac{B \left(2 \log (x)+1+\log (2\pi)-\psi ^{(0)}\left(\frac{3}{2}\right)\right)}{16 \pi }\Big)+O\left((d-3)^1,B^2\right) .
\end{eqnarray} 
It is interesting that (\ref{Diraccurrentd}) works well in two dimensions. Note that since there is no magnetic field in two dimensions,  $B$ should be understood as electric field and $J^{y_1}_{2d}$ should be understood as charge density in (\ref{Diraccurrent2d}). 
Similar to the case of scalar,  the formula (\ref{Diraccurrentd}) does not work well in three dimensions. One can regularize the 3d current in the same way as (\ref{regularize3d}).  We leave a careful study of the 3d current to future work.

\section{Current in a strip}

We have investigated the anomalous current in a half space $0 \le x$. Now let us go on to study the anomalous current in a strip $0 \le x \le L$. For simplicity, we mainly focus on four dimensions in this section. 

\subsection{Complex scalar}

Let us first discuss the complex scalar. The approach is similar to that of sect. 3.1.  Applying (\ref{Gcper}),  we get Green's function
\begin{eqnarray}\label{GscalarStrip}
G(x', x'')=G_0(x', x'')+2i  B \int_0^{L}dx \int_{-\infty}^{\infty}dtdy^{d-2} G_0(x', x) x \overleftarrow{\partial}_{y_1}G_0(x,x'')+O(B^2), 
\end{eqnarray} 
with the background one 
\begin{eqnarray}\label{G0scalarstrip}
&&G_0(x', x'')=\frac{\Gamma \left(\frac{d}{2}-1\right)}{4 \pi ^{d/2}} \sum_{m=-\infty}^{\infty}\Big( \frac{i}{[(x'-x''+2L m)^2+(y'_a-y''_a)^2-(t'-t'')^2]^{\frac{d-2}{2}}} \nonumber\\
&& \ \ \ \ \ \ \ \ \ \ \ \  \ \ \ \ \ \ \ \ \ \ \ \  \ \ \ \ \ \ \ \  +\chi \frac{i}{[(x'+x''+2L m)^2+(y'_a-y''_a)^2-(t'-t'')^2]^{\frac{d-2}{2}}} \Big),
\end{eqnarray} 
which can be derived by the mirror method. 
Note that there are infinite images for two parallel mirrors, and each image corresponds to one $m$ of (\ref{G0scalarstrip}). The reference Green's functions are still given by (\ref{barGscalar},\ref{barG0scalar}).  To simplify the deduction, we rewrite the reference Green's function (\ref{barGscalar}) as
\begin{eqnarray}\label{barGscalarStrip}
&&\bar{G}(x', x'')=\bar{G}_0(x', x'')+O(B^2)\nonumber\\
&&\ \ \ +2i  B \sum_{m=-\infty}^{\infty}\int_0^{L}dx \int_{-\infty}^{\infty}dtdy^{d-2} \bar{G}_0(x', x+m L) (x+m L) \overleftarrow{\partial}_{y_1} \bar{G}_0(x+m L,x''),
\end{eqnarray} 
where $\bar{G}_0$ is given by (\ref{barG0scalar}).  Now let us focus on four dimensions $d=4$.  Substituting (\ref{GscalarStrip},\ref{G0scalarstrip}, \ref{barGscalarStrip}) into (\ref{renoJ}) and performing Wick rotation, we get
\begin{eqnarray}\label{ScalarStripIntegral1}
J_{y_1}=\sum_{\bar{m}\ne 0} \frac{Bx}{8 \pi ^2 \bar{m}^2 L^2}+\sum_{m=-\infty}^{\infty} \frac{ \chi  B x}{8 \pi ^2 (L m+x)^2}+B \int_{0}^L dx' \int_0^{\infty} dr \left( g_1+ g_2 \right)+O(B^2),
\end{eqnarray} 
where
\begin{eqnarray}\label{Stripg1}
g_1=\sum_{m=-\infty}^{\infty}\frac{r^2 \left(L m+x'\right) \left(3 \left(L m+x'-x\right)^2+r^2\right)}{3 \pi ^3 \left(\left(L m+x'-x\right)^2+r^2\right)^4}
\end{eqnarray} 
\begin{eqnarray}\label{Stripg2}
&&g_2=\frac{r^2 x'}{3 \pi ^3}\sum_{m_1,m_2=-\infty}^{\infty} \Big[-2 r^2 \left(\frac{\chi }{\left(A_1^2+r^2\right)^2}+\frac{1}{\left(A_2^2+r^2\right)^2}\right) \left(\frac{\chi }{\left(B_1^2+r^2\right)^2}+\frac{1}{\left(B_2^2+r^2\right)^2}\right)\nonumber\\
&&+\left(4 r^2 \left(\frac{\chi }{\left(A_1^2+r^2\right)^3}+\frac{1}{\left(A_2^2+r^2\right)^3}\right)-\frac{3 \chi }{\left(A_1^2+r^2\right)^2}-\frac{3}{\left(A_2^2+r^2\right)^2}\right) \left(\frac{\chi }{B_1^2+r^2}+\frac{1}{B_2^2+r^2}\right) \Big]\nonumber\\
\end{eqnarray} 
with $A_1=2 L m_1+x+x'$,  $A_2=2 L m_1+x-x'$,   $B_1=2 L m_2+x+x'$ and  $B_2=2 L m_2-x+x'$.

The first two terms of (\ref{ScalarStripIntegral1}) are due to the leading term of Green's function $(G_0-\bar{G}_0)$. After the sum, they yield
\begin{eqnarray}\label{Stripj1}
j_1=\frac{B x \left(3 \chi  \csc ^2\left(\frac{\pi  x}{L}\right)+1\right)}{24 L^2}.
\end{eqnarray} 
Let us go on to consider the integral of $g_1$. 
Performing the r integral , we get
\begin{eqnarray}\label{ScalarStripIntegral2}
\int_{0}^L dx' \int_0^{\infty} dr g_1=  \sum_{m=-\infty}^{\infty}\int_{0}^L dx' \frac{ x'+L m}{24 \pi ^2 |x'+L m-x|^3},
\end{eqnarray} 
where $|\ |$ denotes the absolute value. Thus one needs to discuss the cases $m\ge1, \ m\le-1$ and $m=0$, respectively.  After some calculations, we derive the current for $m\ne 0$
\begin{eqnarray}\label{Stripj2}
j_2=-\frac{B \left(L^2-4 L x+2 x^2\right)}{48 \pi ^2 x (L-x)^2}.
\end{eqnarray} 
As for the case $m=0$, the integral (\ref{ScalarStripIntegral2}) is divergent. However, the divergence can be canceled by the integral of $g_2$ with $m_1=m_2=0$.  Combining together the integrals of $g_1$ with $m=0$ and $g_2$ with $m_1=m_2=0$, we get a finite current
\begin{eqnarray}\label{Stripj3}
j_3=-\frac{B \left(L^3 (6 \chi +1)+8 L^2 \chi  x-2 L \chi  x^2-4 \chi  x^3\right)}{48 \pi ^2 L x (L+x)^2}.
\end{eqnarray} 
Now let us turn to the most complicated parts, the contributions from $g_2$ (\ref{Stripg2}). After the radial integration, we have
\begin{eqnarray}\label{ScalarStripIntegral3}
\int_0^{\infty} dr g_2= \frac{-x'}{3\pi^2} \sum_{m_1, m_2=-\infty}^{\infty} \left(\frac{1}{(|A_1|+|B_1|)^3}+\frac{1}{(|A_2|+|B_2|)^3}+\frac{\chi }{(|A_1|+|B_2|)^3}+\frac{\chi }{(|A_2|+|B_1|)^3}\right)
\end{eqnarray} 
One should discuss cases $m_1\ge 1, m_1\le -1, m_1=0$ and $m_2\ge 1, m_2\le -1, m_2=0$, respectively.  We have already considered the case $m_1=m_2=0$ in (\ref{Stripj3}). The other eight cases contribute a current
\begin{eqnarray}\label{Stripj4}
&&j_4=-\frac{B \left(3 \pi  L x \left(L^2-x^2\right)^2 \cot \left(\frac{\pi  x}{L}\right)+\pi ^2 L x^5+\pi ^2 x^6\right)}{72 \left(\pi ^2 L^2 x (L-x)^2 (L+x)^2\right)}\nonumber\\
&&-\frac{B \left(-3 L^6+\left(6+\pi ^2\right) L^5 x+\left(6+\pi ^2\right) L^4 x^2-2 \pi ^2 L^3 x^3-\left(3+2 \pi ^2\right) L^2 x^4\right)}{72 \left(\pi ^2 L^2 x (L-x)^2 (L+x)^2\right)}\nonumber\\
&&+\frac{B \chi  \left(L \left(L (3 L-2 x)-\pi ^2 x (L+x) \csc ^2\left(\frac{\pi  x}{L}\right)\right)-\pi  x \cot \left(\frac{\pi  x}{L}\right) \left(L^2-\pi ^2 x (L-x) \csc ^2\left(\frac{\pi  x}{L}\right)\right)\right)}{24 \pi ^2 L^3 x}.\nonumber\\
\end{eqnarray} 
In the above derivations, we first do the $x'$ integral and then the sums for $m_1$ and $m_2$.  Fortunately, the sum takes a special form $ \sum_{m_1, m_2=1}^{\infty} f(m_1+m_2)$, which can be transformed into only one sum $ \sum_{m=2}^{\infty} (m-1)f(m)$ .  From (\ref{Stripj1},\ref{Stripj2},\ref{Stripj3},\ref{Stripj4}), we finally obtain the anomalous current for complex scalars in a strip
\begin{eqnarray}\label{ScalarStripCurrent}
&&J^{y_1}=B\frac{3 \cot \left(\frac{\pi  x}{L}\right) \left(L^2 (\chi +1)+\pi ^2 \chi  x (x-L) \csc ^2\left(\frac{\pi  x}{L}\right)\right)+\pi  L (L-2 x) \left(3 \chi  \csc ^2\left(\frac{\pi  x}{L}\right)+1\right)}{72 \pi  L^3}\nonumber\\
&&\ \ \ \ \ \ +O(B^2)
\end{eqnarray} 
Note that we have $J^{y_1}=-J_{y_1}=-(j_1+j_2+j_3+j_4)$ and the other components of currents vanish. 
%\begin{eqnarray}\label{stripcurrentScalar}
%J^{y_1}=\begin{cases}\frac{B}{72L^3}\Big[L (L-2 x)-3 \left(L (L-2 x)+\pi  x (x-L) \cot \left(\frac{\pi  x}{L}\right)\right) \csc ^2\left(\frac{\pi  x}{L}\right)\Big],
%\ \ \ \ \ \ \ \ \ \ \ \ \ \ \  \ \ \ \ \ \ \text{DBC}\\
%\frac{B}{72 \pi  L^3} \Big[3 \cot \left(\frac{\pi  x}{L}\right) \left(2 L^2+\pi ^2 x (x-L) \csc ^2\left(\frac{\pi  x}{L}\right)\right)+\pi  L (L-2 x) \left(3 \csc ^2\left(\frac{\pi  x}{L}\right)+1\right)\Big],
%\ \ \ \text{NBC} .
%\end{cases}
%\end{eqnarray}
Since the magnetic field is a constant in the strip, the current is expected to be antisymmetric, i.e., $J^{y_1}(x)=-J^{y_1}(L-x)$. This is indeed the case of (\ref{ScalarStripCurrent}), which is a strong support of our results. Besides, the above currents have the correct limit (\ref{currentscalar}) near the boundary
\begin{eqnarray}\label{stripcurrentScalarlimit}
J^{y_1}\sim \begin{cases} \frac{B}{24 \pi ^2 x},
\ \ \ \ \ \ \ \ \frac{x}{L} \to 0\\
\frac{B}{24 \pi ^2 (x-L)},
\ \ \ \frac{L-x}{L} \to 0 ,
\end{cases}
\end{eqnarray}
for both DBC $\chi=-1$ and NBC $\chi=1$. This is also a test of our calculations. For the convenience of readers, let us draw a figure for  the anomalous current in a strip. Without loss of generality, we set $B=L=1$.  As showed in figure 1, the current of NBC is larger than the one of DBC. 
%vbird
\begin{figure}[t]
\centering
\includegraphics[width=10cm]{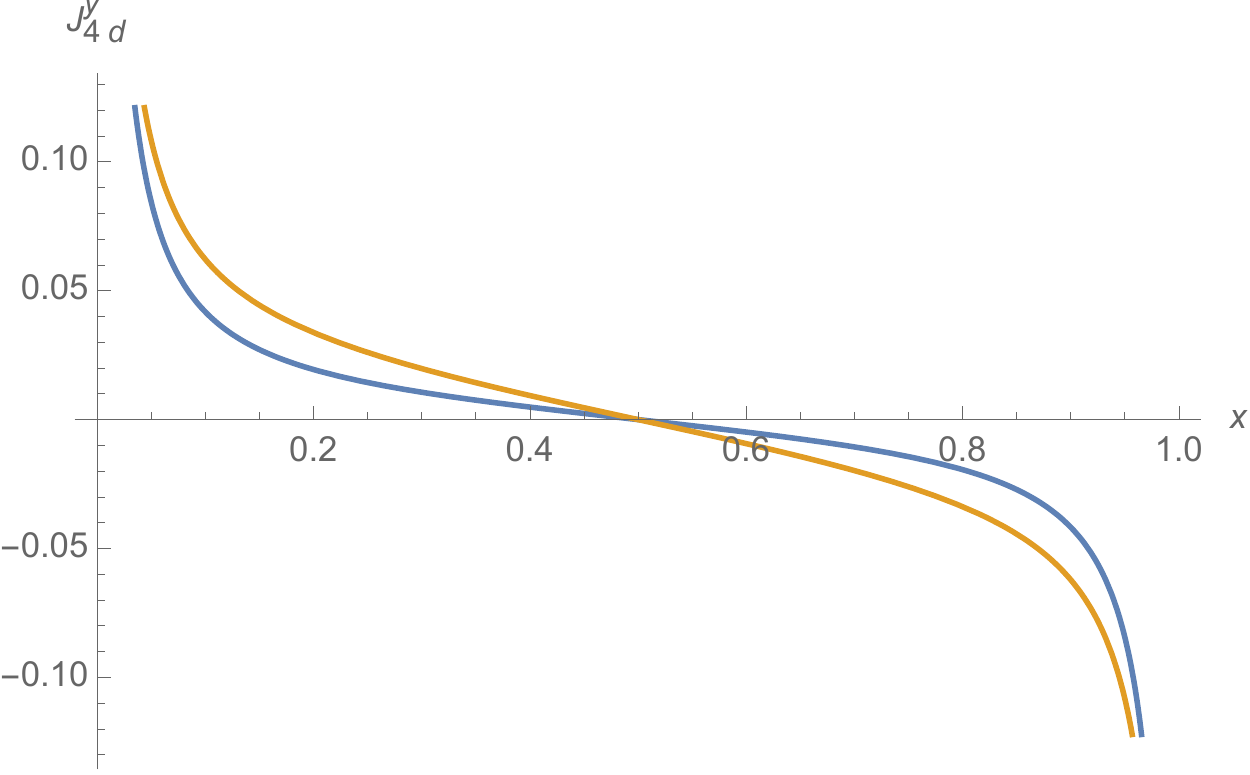}
\caption{Anomalous current for scalars in a strip for DBC (blue line) and NBC (yellow line) in four dimensions.}
\end{figure}

It is straightforward to generalize the above discussions to higher dimensions. Unfortunately, we do not find a general formula. Instead, we calculate them case by case. Please see below for some examples
\begin{eqnarray}\label{ScalarStripCurrent5d}
&&J^{y_1}_{5d}= \frac{-B}{1024 \pi ^2 L^4 x^4} \Big[   6 L^4 \chi  (L-x) (L+3 x)-24 L x^4 \zeta (3) (L-2 x)+2 \pi ^2 x^4 \left(L x-2 \pi ^2 \chi  x^2\right) \csc ^2\left(\frac{\pi  x}{L}\right)  \nonumber\\
&&\ +2 \pi ^2 x^4 \csc ^2\left(\frac{\pi  x}{L}\right) \left(L^2 \left(2 \pi ^2 \chi +\chi -2\right)+3 \pi ^2 \chi  \left(x^2-L^2\right) \csc ^2\left(\frac{\pi  x}{L}\right)+2 \pi  L (\chi +1) x \cot \left(\frac{\pi  x}{L}\right)\right) \nonumber\\
&&\ +x^4 \left(2 L (L (\chi +4)-x) \psi ^{(1)}\left(1-\frac{x}{L}\right)-2 L (3 L \chi +x) \psi ^{(1)}\left(\frac{x}{L}\right)+2 L (3 L \chi -5 \chi  x+x) \psi ^{(2)}\left(\frac{x}{L}\right)\right) \nonumber\\
&&\ +x^4 \left(\chi  (L-x)^2 \psi ^{(3)}\left(1-\frac{x}{L}\right)+\chi  (L+3 x) (L-x) \psi ^{(3)}\left(\frac{L+x}{L}\right)+2 L (3 L \chi -7 \chi  x-x) \psi ^{(2)}\left(1-\frac{x}{L}\right)\right) \Big],\nonumber\\
\end{eqnarray} 
\begin{eqnarray}\label{ScalarStripCurrent6d}
&&J^{y_1}_{6d}= \frac{-B}{28800 \pi ^3 L^5 x^4}\Big[ 360 L^5 \chi  (L-x)-8 \pi ^4 L x^4 (L-2 x)
\nonumber\\
&&\  +30 \pi ^3 L x^4 \csc ^2\left(\frac{\pi  x}{L}\right) \left((3 L-x) \cot \left(\frac{\pi  x}{L}\right)-4 \pi  \chi  (L-2 x) \left(3 \csc ^2\left(\frac{\pi  x}{L}\right)-2\right)\right)\nonumber\\
&&\ +15 L x^4 (-2 L \chi -5 L+x) \psi ^{(2)}\left(1-\frac{x}{L}\right)+15 L x^4 (L (2 \chi +5)-x) \psi ^{(2)}\left(\frac{x}{L}\right)\nonumber\\
&&\  +15 \chi  x^5 (L-x) \left(\psi ^{(4)}\left(1-\frac{x}{L}\right)-\psi ^{(4)}\left(\frac{L+x}{L}\right)\right) \Big],
\end{eqnarray} 
where $\psi $ denotes the PolyGamma function.  It is interesting that DBC $\chi=-1$ and NBC $\chi=1$ yield different directions of currents in dimensions higher than four.  See Figure 2 for example. 

%vbird
\begin{figure}[t]
\centering
\includegraphics[width=7.5cm]{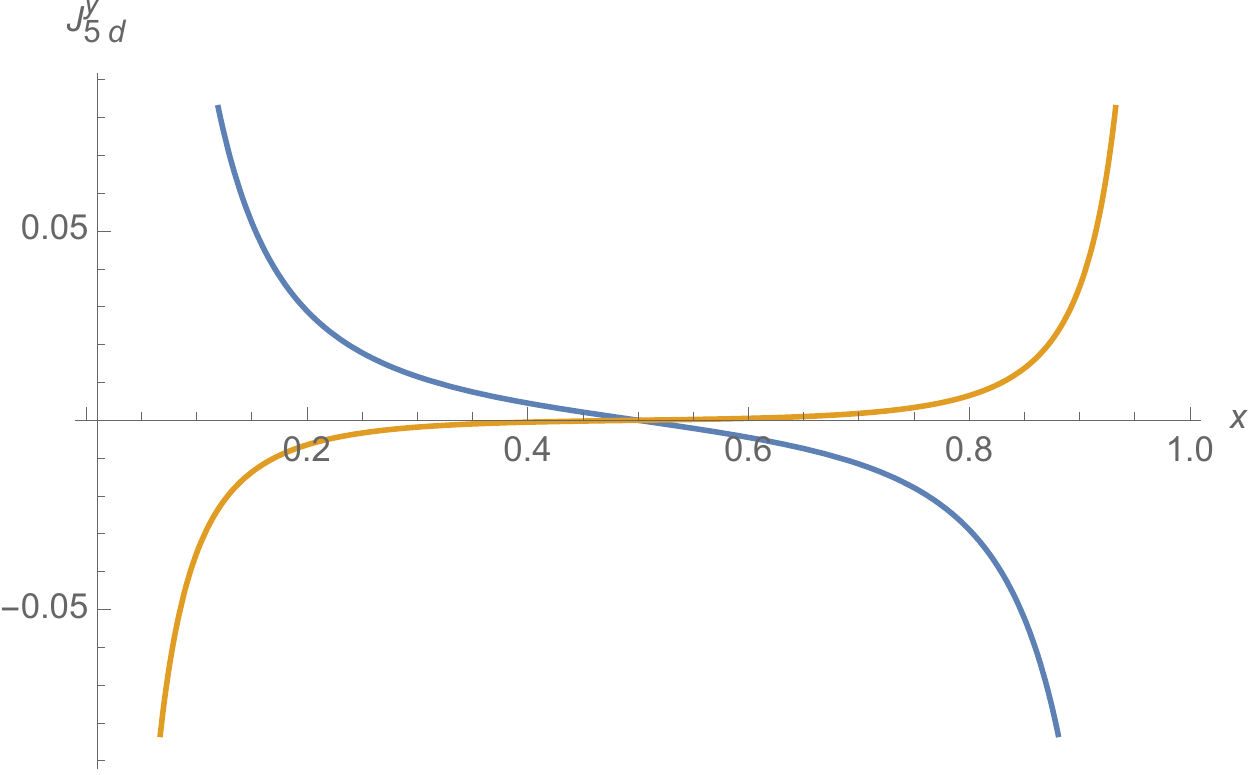} \includegraphics[width=7.5cm]{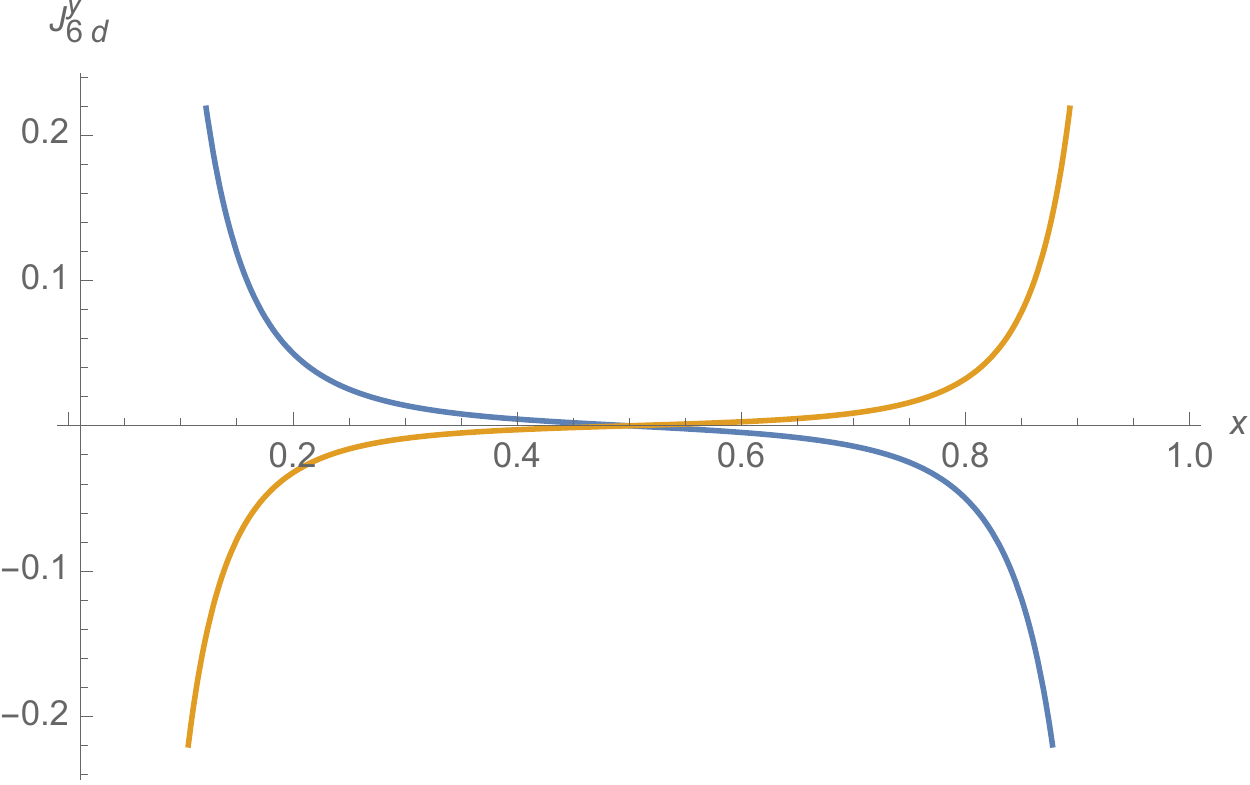}
\caption{The left (right) figure denotes 5d (6d) anomalous current for complex scalars in a strip. Blue line is for DBC and yellow line is for NBC. }
\end{figure}

\subsection{Dirac field}

In this subsection, let us investigate the anomalous current of Dirac field in a strip. For simplicity, we mainly focus on four dimensions. By applying the key formula (\ref{Scper}), we get Green's function for Dirac field
\begin{eqnarray}\label{SDiracStrip}
S(x', x'')=S_0(x', x'')- B \int_0^{L}dx \int_{-\infty}^{\infty}dtdy^{d-2} S_0(x', x) \gamma^2 x S_0(x,x'')+O(B^2),
\end{eqnarray} 
where 
\begin{eqnarray}\label{S0DiracStrip}
&&S_0(x', x'')= \sum_{m=-\infty}^{\infty}\frac{ (d-2) \Gamma \left(\frac{d-2}{2}\right)}{4\pi ^{\frac{d}{2}}} \Big( \frac{\gamma^0(t'-t'') -\gamma^1 (2mL+x'-x'')-\gamma^a (y'_a-y''_a)}{[(2mL+x'-x'')^2+(y'_a-y''_a)^2-(t'-t'')^2]^{\frac{d}{2}}}\nonumber\\
&&\ \ \ \ \ \ \ \ \  \ \ \ \ \ \ \ \ \ \ \ \ \ \ \ \ \ \ \ \ \ \ \ \ +\chi. \frac{\gamma^0(t'-t'') -\gamma^1 (-2mL-x'-x'')-\gamma^a (y'_a-y''_a)}{[(2mL+x'+x'')^2+(y'_a-y''_a)^2-(t'-t'')^2]^{\frac{d}{2}}}\Big)
\end{eqnarray} 
Substituting (\ref{SDiracStrip},\ref{S0DiracStrip}, \ref{barS0Dirac}) into (\ref{regCurrentDirac}), we get the renormalized current
\begin{eqnarray}\label{DiracStripIntegral1}
J^{y_1}=B \int_{0}^L dx' \int_0^{\infty} dr \left( h_1+ h_2 \right)+O(B^2),
\end{eqnarray} 
where 
\begin{eqnarray}\label{DiracStriph1}
h_1=-\sum_{m=-\infty}^{\infty}\frac{4 r^2 \left(L m+x'\right) \left(3 \left(L m+x'-x\right)^2+r^2\right)}{3 \pi ^3 \left(\left(L m+x'-x\right)^2+r^2\right)^4},
\end{eqnarray} 
\begin{eqnarray}\label{DiracStriph2}
h_2=\sum_{m_1,m_2=-\infty}^{\infty}\frac{4 r^2 x'}{3 \pi ^3}\left(\frac{3 A_1 B_1+r^2}{\left(A_1^2+r^2\right){}^2 \left(B_1^2+r^2\right){}^2}+\frac{r^2-3 A_2 B_2}{\left(A_2^2+r^2\right){}^2 \left(B_2^2+r^2\right){}^2} \right)
\end{eqnarray} 
Recall that $A_1=2 L m_1+x+x'$,  $A_2=2 L m_1+x-x'$,   $B_1=2 L m_2+x+x'$ and  $B_2=2 L m_2-x+x'$.
Following the approaches of sect. 3.1, we can derive the current. Since the calculations are similar to those of sect. 3.1, we do not repeat them here. We obtain 
\begin{eqnarray}\label{currentstripDirac}
J^{\mu}=\delta^{\mu}_{y_1}\frac{B \left(6 L \cot \left(\frac{\pi  x}{L}\right)-\pi  L+2 \pi  x\right)}{36 \pi  L^2}+O(B^2),
\end{eqnarray} 
which obeys the relation $J^{y_1}(x)=-J^{y_1}(L-x)$ and reduces to the current (\ref{Diraccurrent4d}) in a half space in the near-boundary limit
\begin{eqnarray}\label{stripcurrentScalarlimit}
J^{y_1}\sim \begin{cases} 
\frac{B}{6 \pi^2 x},
\ \ \ \ \ \ \ \ \frac{x}{L} \to 0\\
\frac{B}{6\pi^2 (x-L)},
\ \ \ \frac{L-x}{L} \to 0. 
\end{cases}
\end{eqnarray}
Similar to the case in a half space, the current (\ref{currentstripDirac}) in a strip is independent of BCs (\ref{ChiralBC}) too. 
To compare with the currents of scalars, let us draw a figure. From Figure 3, we notice that the current of Dirac field is always larger than those of scalars. 
%vbird
\begin{figure}[t]
\centering
\includegraphics[width=10cm]{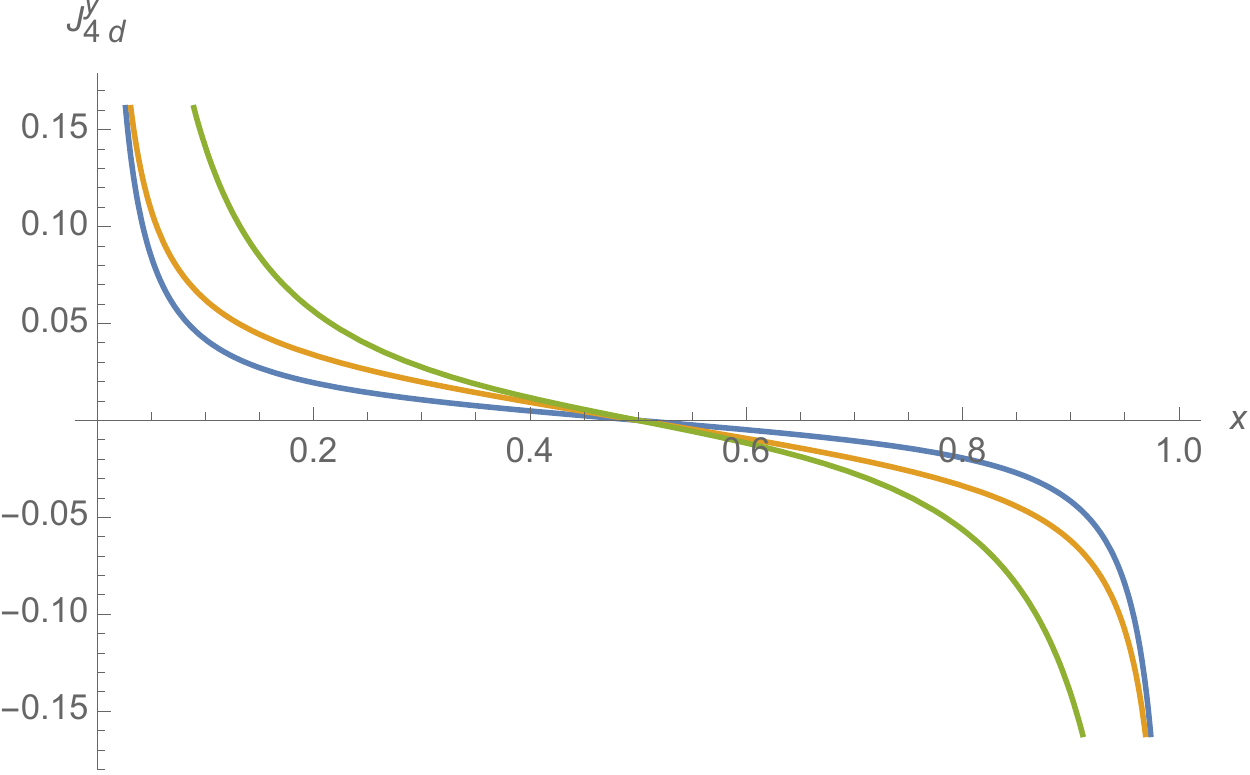}
\caption{Current in a strip for Dirac field (green line)  and scalars (blue line for DBC and yellow line for NBC) in four dimensions.}
\end{figure}

The generalizations to higher dimensions are straightforward.  Following the above approach, we get anomalous currents for Dirac fields
\begin{eqnarray}\label{DiracStripCurrent5d}
J^{y_1}_{5d}=\frac{3}{128 \pi ^2}\left(\frac{-2 \zeta (3) (L-2 x)-L \psi ^{(1)}\left(2-\frac{x}{L}\right)+L \psi ^{(1)}\left(\frac{x}{L}\right)}{L^3}-\frac{1}{(L-x)^2}\right)B+O(B^2),
\end{eqnarray} 
in five dimensions and
\begin{eqnarray}\label{DiracStripCurrent6d}
J^{y_1}_{6d}= \frac{30 L \cot \left(\frac{\pi  x}{L}\right) \csc ^2\left(\frac{\pi  x}{L}\right)-\pi  L+2 \pi  x}{900 L^4} B+O(B^2).
\end{eqnarray} 
in six dimensions. Similar to the 4d case, currents of Dirac fields are independent of BCs (\ref{ChiralBC}) in higher dimensions. Besides, near the boundary, the currents become larger as the dimensions increase. See Figure 4 for example.  
%vbird
\begin{figure}[t]
\centering
\includegraphics[width=10cm]{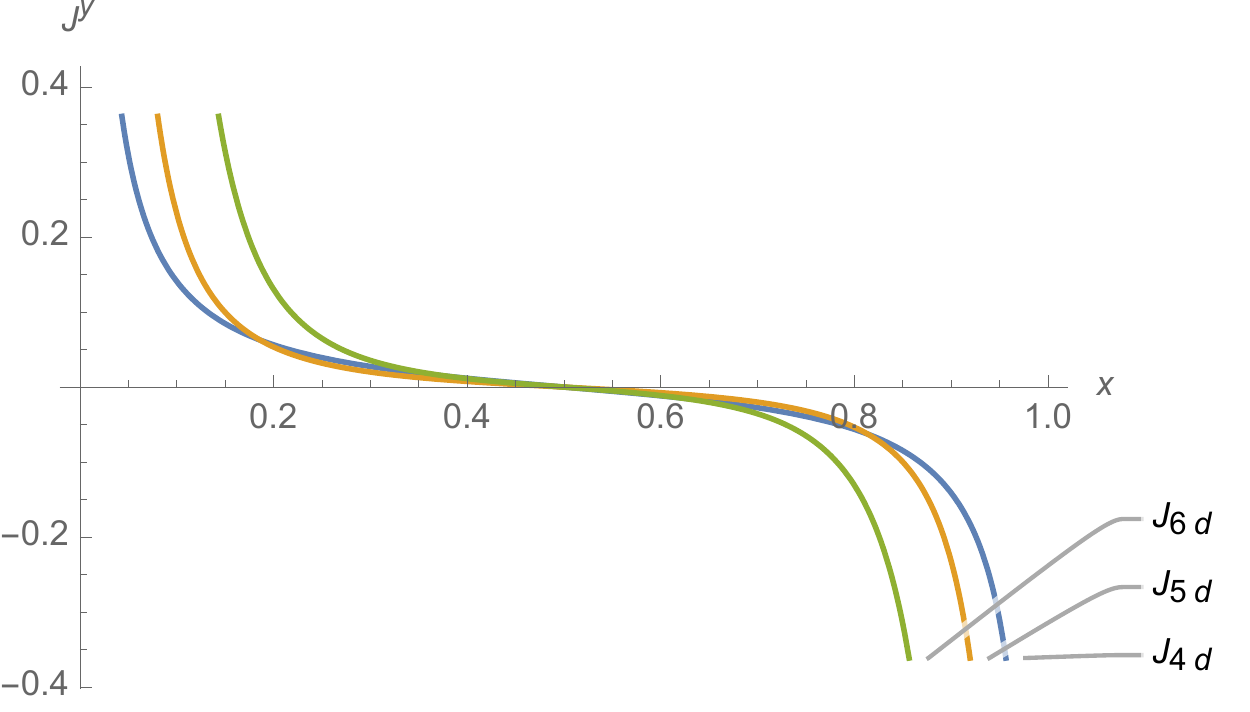}
\caption{ Anomalous currents of Dirac fields in a strip in 4,5,6 dimensions.   }
\end{figure}

In summary, we have obtained the anomalous currents for complex scalars and Dirac fields in a strip. Our results are exact in the size of strip $L$. In other words, we have got the anomalous currents beyond the near-boundary regions of \cite{Chu:2018ksb,Chu:2018ntx}. 

\section{Non-perturbative results}

In previous sections, we focus on the anomalous current at the linear order of $O(B)$. In this section, we try to derive the current exactly in the magnetic field $B$. For simplicity, we focus on the complex scalar in a half space with a constant magnetic field $A_{\mu}=(0,0,Bx, 0,...,0)$.  We get a formal expression of currents which can be evaluated numerically. 

Recall that Green's function obeys EOM
\begin{eqnarray}\label{GreenEOMexact}
[D^{\mu}D_{\mu}+m^2 ]G(x,x')=\delta^{(d)}(x,x'),
\end{eqnarray}
where $m$ is the mass. Performing Fourier transform for the tangential coordinates, 
\begin{eqnarray}\label{Fouriertransform}
G=\int \frac{dk_{\parallel}^{d-1}}{(2\pi)^{d-1}}\tilde{G}(k) e^{-i k_{\parallel}.\Delta x_{\parallel}}
\end{eqnarray}
we can rewrite (\ref{GreenEOMexact}) as
\begin{eqnarray}\label{GreenEOMexact1}
[-\partial_x^2+(m^2+k^2_{\parallel}) -2 B x k_{1}+ B^2 x^2]\tilde{G}=\delta(x-x').
\end{eqnarray}
Here $k_1=k_{y_1}$, $k_{\parallel}.\Delta x_{\parallel}=k^0 (t-t')-k^a (y_a-y_a')$ and $k^2_{\parallel}=k_a^2-k_0^2$. We split Green's function $\tilde{G}$ into the one in free space and a correction due to the boundary
\begin{eqnarray}\label{Gk}
\tilde{G}=G_{free}+ G_{bdy},
\end{eqnarray}
where $G_{free}$ is given by \cite{JohnP,FalKovskii}
\begin{eqnarray}\label{Gfree}
G_{free}=\begin{cases}\sqrt{\frac{1}{4 \pi B}}  \Gamma (\lambda_k ) D_{-\lambda_k }\left(\sqrt{2} \left(\bar{x}-\bar{k}_1\right)\right) D_{-\lambda_k }\left(\sqrt{2} \left(\bar{k}_1-\bar{x'}\right)\right)
,
\ \ \ \ \ \ \ \ x> x',\\
\sqrt{\frac{1}{4 \pi B}}  \Gamma (\lambda_k ) D_{-\lambda_k }\left(\sqrt{2} \left(\bar{k}_1-\bar{x}\right)\right) D_{-\lambda_k }\left(\sqrt{2} \left(\bar{x'}-\bar{k}_1\right)\right)
,
\ \ \ \ \ \ \ \ x<x'.
\end{cases}
\end{eqnarray}
Here $D$ denotes the parabolic cylinder function, $\lambda_k=(B+m^2+k^2_{\parallel}-k_1^2)/(2 B)$, $\bar{k}_1=k_1/\sqrt{B}$ and $\bar{x}=\sqrt{B}x$.  Note that our convention of Fourier transform (\ref{Fouriertransform}) is different from that of \cite{JohnP}. As a result, $G_{free}$ (\ref{Gfree}) differs by a factor $2\pi$ from the one of \cite{JohnP}. Imposing BCs (\ref{GreenDBC}, \ref{GreenNBC}), we solve the corrections to Green's function
\begin{eqnarray}\label{GbdyDBC}
G_{bdy}=\frac{-\Gamma \left(\lambda _k\right) D_{-\lambda _k}\left(\sqrt{2} \bar{k}_1\right) }{2 \pi ^{1/2} \sqrt{B} D_{-\lambda _k}\left(-\sqrt{2} \bar{k}_1\right)} D_{-\lambda _k}\left(\sqrt{2} \left( \bar{x}-\bar{k}_1\right)\right)D_{-\lambda _k}\left(\sqrt{2} \left(\bar{x'}-\bar{k}_1\right)\right)
\end{eqnarray}
for DBC and 
\begin{eqnarray}\label{GbdyNBC}
G_{bdy}=\frac{\Gamma \left(\lambda _k\right) \left(\sqrt{2} D_{1-\lambda _k}\left(\sqrt{2} \bar{k}_1\right)- \bar{k}_1 D_{-\lambda _k}\left(\sqrt{2} \bar{k}_1\right)\right)}{2 \pi ^{1/2} \sqrt{B} \left(\sqrt{2} D_{1-\lambda _k}\left(-\sqrt{2} \bar{k}_1\right)+\bar{k}_1 D_{-\lambda _k}\left(-\sqrt{2} \bar{k}_1\right)\right)} D_{-\lambda _k}\left(\sqrt{2} \left( \bar{x}-\bar{k}_1\right)\right)D_{-\lambda _k}\left(\sqrt{2} \left(\bar{x'}-\bar{k}_1\right)\right)\nonumber\\
\end{eqnarray}
for NBC.  (\ref{GbdyDBC}) for DBC agrees with \cite{JohnP} and (\ref{GbdyNBC}) for NBC is a new result.

Now we are ready to derive the anomalous current. Substituting (\ref{Fouriertransform},\ref{GbdyDBC},\ref{GbdyNBC}) into (\ref{renoJ}) and performing the Wick rotation $k^0\to i k_E$ \cite{Peskin:1995ev}, we get the renormalized current
\begin{eqnarray}\label{CurrentDBC}
J^{y_1}=\int_{-\infty}^{\infty} dp^{d-2}dk_1\frac{\left(\bar{x}-\bar{k}_1\right) \Gamma \left(\lambda _p\right) D_{-\lambda _p}\left(\sqrt{2} \bar{k}_1\right)}{2^{d-1} \pi ^{d-\frac{1}{2}} D_{-\lambda _p}\left(-\sqrt{2} \bar{k}_1\right)}D_{-\lambda _p}\left(\sqrt{2} \left(\bar{x}-\bar{k}_1\right)\right){}^2
\end{eqnarray}
for DBC and 
\begin{eqnarray}\label{CurrentNBC}
&&J^{y_1}=-\int_{-\infty}^{\infty}dp^{d-2}dk_1 \frac{(\bar{x}-\bar{k}_1) \Gamma \left(\lambda _p\right) \left(\sqrt{2} D_{1-\lambda _p}\left(\sqrt{2} \bar{k}_1\right)-\bar{k}_1 D_{-\lambda _p}\left(\sqrt{2} \bar{k}_1\right)\right)}{2^{d-1} \pi ^{d-\frac{1}{2}} \left(\sqrt{2} D_{1-\lambda _p}\left(-\sqrt{2} \bar{k}_1\right)+\bar{k}_1 D_{-\lambda _p}\left(-\sqrt{2} \bar{k}_1\right)\right)}\nonumber\\
&&\ \ \ \ \ \ \ \ \ \ \ \ \ \times D_{-\lambda _p}\left(\sqrt{2} \left(\bar{x}-\bar{k}_1\right)\right){}^2
\end{eqnarray}
for NBC. Recall that $J^{y_1}=-J_{y_1}$, $\lambda_p=(B+m^2+p^2)/(2 B)$, $\bar{k}_1=k_1/\sqrt{B}$ and $\bar{x}=\sqrt{B}x$.  In principle, the formal expressions (\ref{CurrentDBC},\ref{CurrentNBC}) can be evaluated numerically.  Let us take $d=2$ as an example, where the $p$ integral disappears, which can simplify calculations greatly. Note that, for $d=2$, $B$ of (\ref{CurrentDBC},\ref{CurrentNBC}) should be understood as the electric field, and $J_{y_1}$ should be  understood as the charge density. The results are shown in Figure 5, which implies that the mass suppresses the current and the current approaches zero far aways from the boundary.  These are the expected behaviors. Note that the numerical integration does not work well near the boundary $x\sim 0$. In the near-boundary region, we can obtain the anomalous current by using methods of sect. 3.1. See (\ref{scalarcurrent2d})(\ref{regularize2d}) for example. 
%vbird
\begin{figure}[t]
\centering
\includegraphics[width=7.5cm]{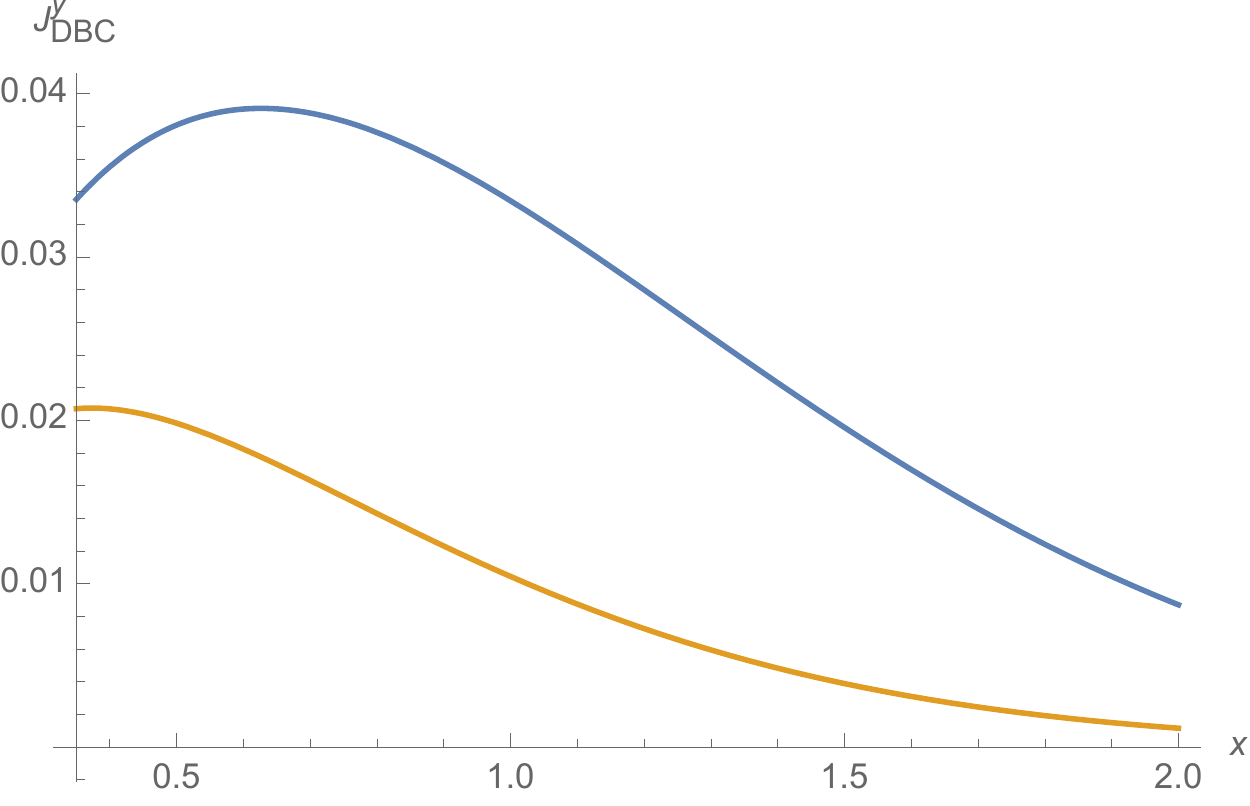}
\includegraphics[width=7.5cm]{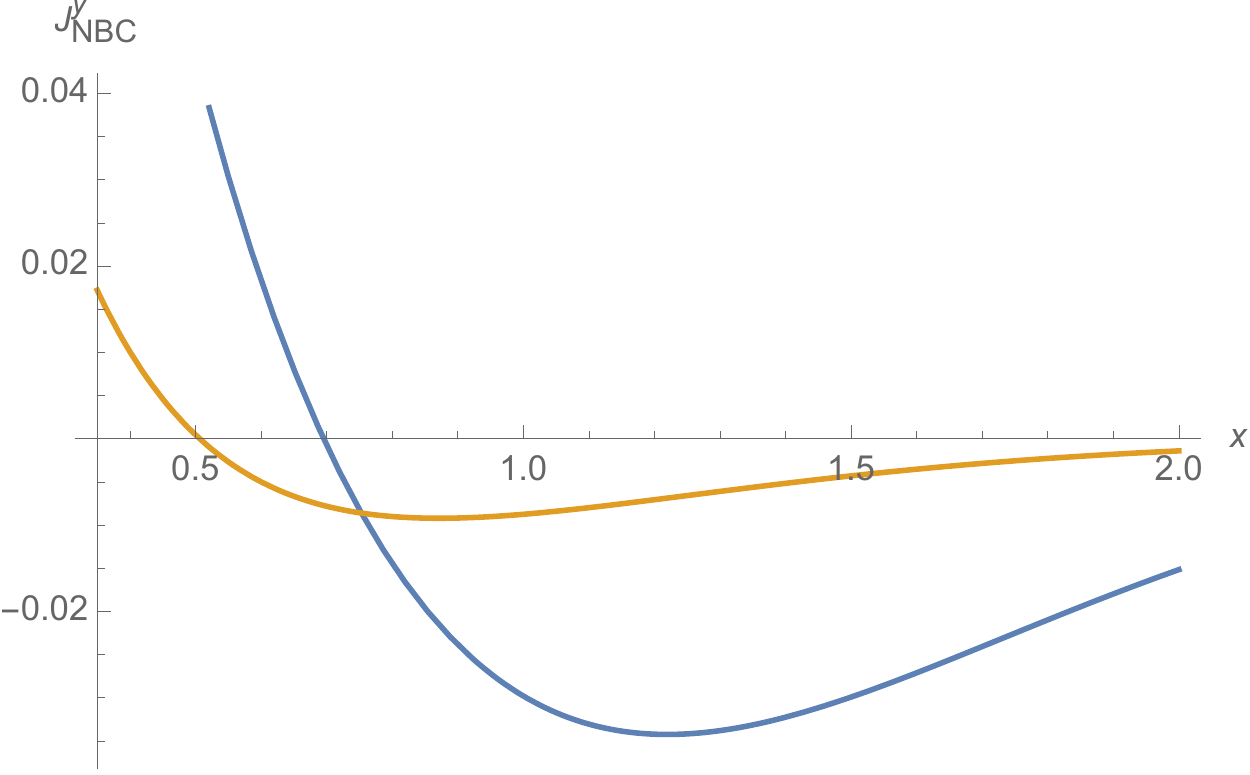}
\caption{The left figure is for 2d current of DBC and the right figure is for 2d current of NBC. Blue line denotes zero mass m=0 and yellow line denotes finite mass m=1. }
\end{figure}

\section{Conclusions and Discussions}

In this paper, we have investigated the anomalous current for free theories in the spacetime with boundaries.  In a half space, we get the anomalous current at the linear order of magnetic fields in general dimensions. The currents of scalars agree with those in the literature. And the currents for Dirac field are new. Our results work well in the region close to the boundary, i.e., $B x^2 <1$. We also obtain the anomalous currents in a strip. The currents are of the linear order of magnetic field $B$ and exact in the size of strip $L$. This means that our results apply to the full region 
of a strip, as long as the magnetic field is small $B L^2<1$. It is remarkable that, unlike the scalar and holographic BCFT, the anomalous currents of free Dirac fields are independent of boundary conditions in general dimensions. It should be mentioned that, although we focus on the constant magnetic field in this paper, our approaches apply to arbitrary electromagnetic fields as well.  Finally, we derive a formal expression of anomalous current for complex scalars in a half space.  The numerical results imply that the mass suppresses the anomalous currents. In this paper, we mainly focus on dimensions higher than three.  It is interesting to study carefully the cases in two and three dimensions. Besides, it is also interesting to study the effect of temperature and the anomalous current with other shapes of boundaries such as cylinders and balls. We hope we could address these problem in future.

\section*{Acknowledgements}
R. X. Miao acknowledges the supports from NSFC grant (No. 11905297) and Guangdong Basic and Applied Basic Research Foundation (No.2020A1515010900).

\appendix


\begin{thebibliography}{00}

\bibitem{review}
  For a review, see for example,
  D.~E.~Kharzeev,
  ``The Chiral Magnetic Effect and Anomaly-Induced Transport,''
  Prog.\ Part.\ Nucl.\ Phys.\  {\bf 75} (2014) 133
  [arXiv:1312.3348 [hep-ph]];
  %%CITATION = doi:10.1016/j.ppnp.2014.01.002;%%
  %151 citations counted in INSPIRE as of 09 Jan 2018
  K.~Landsteiner,
  ``Notes on Anomaly Induced Transport,''
  Acta Phys.\ Polon.\ B {\bf 47} (2016) 2617
  [arXiv:1610.04413 [hep-th]].
  %%CITATION = doi:10.5506/APhysPolB.47.2617;%%

\bibitem{Vilenkin:1995um}
  A.~Vilenkin,
  ``Parity nonconservation and neutrino transport in magnetic fields,''
  Astrophys.\ J.\  {\bf 451} (1995) 700.

  \bibitem{Vilenkin:1980fu}
  A.~Vilenkin,
  ``Equilibrium Parity Violating Current In A Magnetic Field,''
  Phys.\ Rev.\ D {\bf 22} (1980) 3080.
  %%CITATION = doi:10.1103/PhysRevD.22.3080;%%
  %227 citations counted in INSPIRE as of 09 Feb 2018

\bibitem{Giovannini:1997eg}
  M.~Giovannini and M.~E.~Shaposhnikov,
  ``Primordial hypermagnetic fields and triangle anomaly,''
  Phys.\ Rev.\ D {\bf 57} (1998) 2186
  [hep-ph/9710234].
  %%CITATION = doi:10.1103/PhysRevD.57.2186;%%
  %251 citations counted in INSPIRE as of 09 Feb 2018

  \bibitem{alekseev}
  A.Y. Alekseev, V. V. Cheianov, and J. Frohlich,
   Phys. Rev. Lett.{\bf  81} (1998) 3503 [cond-mat/9803346].


%\cite{Fukushima:2012vr}
\bibitem{Fukushima:2012vr}
  K.~Fukushima,
  ``Views of the Chiral Magnetic Effect,''
  Lect.\ Notes Phys.\  {\bf 871} (2013) 241
  [arXiv:1209.5064 [hep-ph]].
  %%CITATION = doi:10.1007/978-3-642-37305-3_9;%%
  %41 citations counted in INSPIRE as of 09 Feb 2018

  %\cite{Kharzeev:2007tn}
\bibitem{Kharzeev:2007tn}
  D.~Kharzeev and A.~Zhitnitsky,
  ``Charge separation induced by P-odd bubbles in QCD matter,''
  Nucl.\ Phys.\ A {\bf 797}, 67 (2007)
  [arXiv:0706.1026 [hep-ph]].
  %%CITATION = doi:10.1016/j.nuclphysa.2007.10.001;%%
  %331 citations counted in INSPIRE as of 09 Feb 2018

\bibitem{Erdmenger:2008rm}
  J.~Erdmenger, M.~Haack, M.~Kaminski and A.~Yarom,
  ``Fluid dynamics of R-charged black holes,''
  JHEP {\bf 0901} (2009) 055
  [arXiv:0809.2488 [hep-th]].
  %%CITATION = doi:10.1088/1126-6708/2009/01/055;%%
  %326 citations counted in INSPIRE as of 09 Feb 2018

\bibitem{Banerjee:2008th}
  N.~Banerjee, J.~Bhattacharya, S.~Bhattacharyya, S.~Dutta, R.~Loganayagam
  and P.~Surowka,
  ``Hydrodynamics from charged black branes,''
  JHEP {\bf 1101} (2011) 094
  [arXiv:0809.2596 [hep-th]].
  %%CITATION = doi:10.1007/JHEP01(2011)094;%%
  %311 citations counted in INSPIRE as of 09 Feb 2018

\bibitem{Son:2009tf}
  D.~T.~Son and P.~Surowka,
  ``Hydrodynamics with Triangle Anomalies,''
  Phys.\ Rev.\ Lett.\  {\bf 103} (2009) 191601
  [arXiv:0906.5044 [hep-th]].

\bibitem{Landsteiner:2011cp}
  K.~Landsteiner, E.~Megias and F.~Pena-Benitez,
  ``Gravitational Anomaly and Transport,''
  Phys.\ Rev.\ Lett.\  {\bf 107} (2011) 021601
  [arXiv:1103.5006 [hep-ph]].

%\cite{Golkar:2012kb}
\bibitem{Golkar:2012kb}
  S.~Golkar and D.~T.~Son,
  ``(Non)-renormalization of the chiral vortical effect coefficient,''
  JHEP {\bf 1502}, 169 (2015)
 % doi:10.1007/JHEP02(2015)169
  [arXiv:1207.5806 [hep-th]].
  %%CITATION = doi:10.1007/JHEP02(2015)169;%%
  %68 citations counted in INSPIRE as of 16 Mar 2018

%\cite{Jensen:2012kj}
\bibitem{Jensen:2012kj}
  K.~Jensen, R.~Loganayagam and A.~Yarom,
  ``Thermodynamics, gravitational anomalies and cones,''
  JHEP {\bf 1302}, 088 (2013)
 % doi:10.1007/JHEP02(2013)088
  [arXiv:1207.5824 [hep-th]].
  %%CITATION = doi:10.1007/JHEP02(2013)088;%%
  %124 citations counted in INSPIRE as of 16 Mar 2018
  
  %\cite{Duff:1993wm}
\bibitem{Duff:1993wm} 
  M.~J.~Duff,
  %``Twenty years of the Weyl anomaly,''
  Class.\ Quant.\ Grav.\  {\bf 11}, 1387 (1994)
%  doi:10.1088/0264-9381/11/6/004
 % [hep-th/9308075].
  %%CITATION = doi:10.1088/0264-9381/11/6/004;%%
  %332 citations counted in INSPIRE as of 02 Nov 2017 


\bibitem{Chernodub:2016lbo}
  M.~N.~Chernodub,
  %``Anomalous Transport Due to the Conformal Anomaly,''
  Phys.\ Rev.\ Lett.\  {\bf 117}, no. 14, 141601 (2016)
 % doi:10.1103/PhysRevLett.117.141601
  [arXiv:1603.07993 [hep-th]].
  %%CITATION = doi:10.1103/PhysRevLett.117.141601;%%
  %9 citations counted in INSPIRE as of 10 Mar 2019

  %\cite{Chernodub:2017jcp}/Users/mac/Downloads/Physical Review Journals - Publicity Instructions for Authors.pdf
\bibitem{Chernodub:2017jcp}
  M.~N.~Chernodub, A.~Cortijo and M.~A.~H.~Vozmediano,
  %``Generation of a Nernst Current from the Conformal Anomaly in Dirac and Weyl Semimetals,''
  Phys.\ Rev.\ Lett.\  {\bf 120}, no. 20, 206601 (2018)
%  doi:10.1103/PhysRevLett.120.206601
  [arXiv:1712.05386 [cond-mat.str-el]].
  %%CITATION = doi:10.1103/PhysRevLett.120.206601;%%
  %10 citations counted in INSPIRE as of 10 Mar 2019

  %\cite{Chu:2018ksb}
\bibitem{Chu:2018ksb}
  C.~S.~Chu and R.~X.~Miao,
  %``Weyl Anomaly Induced Current in Boundary Quantum Field Theories,''
  Phys.\ Rev.\ Lett.\  {\bf 121}, no. 25, 251602 (2018)
  %doi:10.1103/PhysRevLett.121.251602
  [arXiv:1803.03068 [hep-th]].
  %%CITATION = doi:10.1103/PhysRevLett.121.251602;%%
  %10 citations counted in INSPIRE as of 03 Apr 2019

  %\cite{Chu:2018ntx}
\bibitem{Chu:2018ntx}
  C.~S.~Chu and R.~X.~Miao,
  %``Anomalous Transport in Holographic Boundary Conformal Field Theories,''
  JHEP {\bf 1807}, 005 (2018)
 % doi:10.1007/JHEP07(2018)005
  [arXiv:1804.01648 [hep-th]].
  %%CITATION = doi:10.1007/JHEP07(2018)005;%%
  %9 citations counted in INSPIRE as of 03 Apr 2019
  
  %\cite{}


  
  %\cite{Chu:2018fpx,Chu:2019rod,Miao:2017aba,Miao:2018dvm,Chernodub:2018ihb,Chernodub:2019blw,Ambrus:2019khr}
\bibitem{Chu:2018fpx}
C.~Chu and R.~Miao,
%``Boundary String Current & Weyl Anomaly in Six-dimensional Conformal Field Theory,''
JHEP \textbf{07}, 151 (2019)
%doi:10.1007/JHEP07(2019)151
[arXiv:1812.10273 [hep-th]].

  
  %\cite{Chu:2019rod}
\bibitem{Chu:2019rod}
C.~Chu,
%``Weyl Anomaly and Vacuum Magnetization Current of M5‐brane in Background Flux,''
Fortsch.\ Phys.\  \textbf{67}, no.8-9, 1910005 (2019)
%doi:10.1002/prop.201910005
[arXiv:1903.02817 [hep-th]].
  
%\cite{Miao:2017aba}
\bibitem{Miao:2017aba}
R.~Miao and C.~Chu,
%``Universality for Shape Dependence of Casimir Effects from Weyl Anomaly,''
JHEP \textbf{03}, 046 (2018)
%doi:10.1007/JHEP03(2018)046
[arXiv:1706.09652 [hep-th]].  
  
  %\cite{Miao:2018dvm}
\bibitem{Miao:2018dvm}
R.~Miao,
%``Casimir Effect, Weyl Anomaly and Displacement Operator in Boundary Conformal Field Theory,''
JHEP \textbf{07}, 098 (2019)
%doi:10.1007/JHEP07(2019)098
[arXiv:1808.05783 [hep-th]].

  %\cite{Chernodub:2018ihb}
\bibitem{Chernodub:2018ihb}
  M.~N.~Chernodub, V.~A.~Goy and A.~V.~Molochkov,
  %``Conformal magnetic effect at the edge: a numerical study in scalar QED,''
  Phys.\ Lett.\ B {\bf 789}, 556 (2019)
 % doi:10.1016/j.physletb.2019.01.003
  [arXiv:1811.05411 [hep-th]].
  %%CITATION = doi:10.1016/j.physletb.2019.01.003;%%
  %1 citations counted in INSPIRE as of 19 Apr 2019
  
  %\cite{Chernodub:2019blw}
\bibitem{Chernodub:2019blw}
  M.~N.~Chernodub, and M.~A.~H.~Vozmediano,
%``Direct measurement of a beta function and an indirect check of the Schwinger effect near the boundary in Dirac-Weyl semimetals,''
Phys.\ Rev.\ Research.\  \textbf{1}, 032002 (2019)
%doi:10.1103/PhysRevResearch.1.032002
[arXiv:1902.02694 [cond-mat.str-el]].

%\cite{Ambrus:2019khr}
\bibitem{Ambrus:2019khr}
V.~E.~Ambrus and M.~Chernodub,
%``Helical vortical effects, helical waves, and anomalies of Dirac fermions,''
[arXiv:1912.11034 [hep-th]].

%\cite{Zheng:2019xeu}
\bibitem{Zheng:2019xeu}
J.~Zheng, D.~Li, Y.~Zeng and R.~Miao,
%``Anomalous Current Due to Weyl Anomaly for Conformal Field Theory,''
Phys.\ Lett.\ B \textbf{797}, 134844 (2019)
%doi:10.1016/j.physletb.2019.134844
[arXiv:1904.07017 [hep-th]].


%\cite{Chu:2020mwx}
\bibitem{Chu:2020mwx}
C.~Chu and R.~Miao,
``Fermi Condensation induced by Weyl Anomaly,''
[arXiv:2004.05780 [hep-th]].




%\cite{Casimir:1948dh,Plunien:1986ca,Bordag:2001qi}
\bibitem{Casimir:1948dh}
  H.~B.~G.~Casimir,
  %``On the Attraction Between Two Perfectly Conducting Plates,''
  Indag.\ Math.\  {\bf 10} (1948) 261
   [Kon.\ Ned.\ Akad.\ Wetensch.\ Proc.\  {\bf 51} (1948) 793]
   [Front.\ Phys.\  {\bf 65} (1987) 342]
   [Kon.\ Ned.\ Akad.\ Wetensch.\ Proc.\  {\bf 100N3-4} (1997) 61].
  %%CITATION = IMTHB,10,261;%%
  %1153 citations counted in INSPIRE as of 23 Jun 2017    

%\cite{Plunien:1986ca}
\bibitem{Plunien:1986ca}
  G.~Plunien, B.~Muller and W.~Greiner,
  %``The Casimir Effect,''
  Phys.\ Rept.\  {\bf 134} (1986) 87.
 % doi:10.1016/0370-1573(86)90020-7
  %%CITATION = doi:10.1016/0370-1573(86)90020-7;%%
  %447 citations counted in INSPIRE as of 23 Jun 2017

%\cite{Bordag:2001qi}
\bibitem{Bordag:2001qi}
  M.~Bordag, U.~Mohideen and V.~M.~Mostepanenko,
  %``New developments in the Casimir effect,''
  Phys.\ Rept.\  {\bf 353} (2001) 1
 % doi:10.1016/S0370-1573(01)00015-1
  [quant-ph/0106045].
  %%CITATION = doi:10.1016/S0370-1573(01)00015-1;%%
  %768 citations counted in INSPIRE as of 23 Jun 2017


%\cite{Hasan:2010xy}
\bibitem{Hasan:2010xy}
  M.~Z.~Hasan and C.~L.~Kane,
  %``Topological Insulators,''
  Rev.\ Mod.\ Phys.\  {\bf 82}, 3045 (2010)
 % doi:10.1103/RevModPhys.82.3045
 % [arXiv:1002.3895 [cond-mat.mes-hall]].
  %%CITATION = doi:10.1103/RevModPhys.82.3045;%%
  %1156 citations counted in INSPIRE as of 12 Dec 2017
  
  
  %\cite{Cardy:2004hm,McAvity:1993ue}
\bibitem{Cardy:2004hm} 
  J.~L.~Cardy,
  %``Boundary conformal field theory,''
  hep-th/0411189.
  %%CITATION = HEP-TH/0411189;%%
  %57 citations counted in INSPIRE as of 23 Dec 2016
  
  %\cite{McAvity:1993ue}
\bibitem{McAvity:1993ue} 
  D.~M.~McAvity and H.~Osborn,
  %``Energy momentum tensor in conformal field theories near a boundary,''
  Nucl.\ Phys.\ B {\bf 406}, 655 (1993)
  [hep-th/9302068].
  %%CITATION = doi:10.1016/0550-3213(93)90005-A;%%
  %25 citations counted in INSPIRE as of 22 Dec 2016


%\cite{Takayanagi:2011zk}
\bibitem{Takayanagi:2011zk}
  T.~Takayanagi,
  %``Holographic Dual of BCFT,''
  Phys.\ Rev.\ Lett.\  {\bf 107} (2011) 101602
  [arXiv:1105.5165 [hep-th]].
  %%CITATION = doi:10.1103/PhysRevLett.107.101602;%%
  %68 citations counted in INSPIRE as of 14 Dec 2016



%\cite{Jensen:2015swa,Fursaev:2015wpa,Herzog:2015ioa,Herzog:2017kkj,Herzog:2017xha}
\bibitem{Jensen:2015swa} 
  K.~Jensen and A.~O'Bannon,
  %``Constraint on Defect and Boundary Renormalization Group Flows,''
  Phys.\ Rev.\ Lett.\  {\bf 116}, no. 9, 091601 (2016)
  %doi:10.1103/PhysRevLett.116.091601
  [arXiv:1509.02160 [hep-th]].
  %%CITATION = doi:10.1103/PhysRevLett.116.091601;%%
  %16 citations counted in INSPIRE as of 29 Jun 2017



%\cite{Fursaev:2015wpa}
\bibitem{Fursaev:2015wpa} 
  D.~Fursaev,
  %``Conformal anomalies of CFT’s with boundaries,''
  JHEP {\bf 1512}, 112 (2015)
 [arXiv:1510.01427 [hep-th]].
  %%CITATION = doi:10.1007/JHEP12(2015)112;%%
  %7 citations counted in INSPIRE as of 22 Dec 2016 

%c4
%\cite{Herzog:2015ioa}
\bibitem{Herzog:2015ioa} 
  C.~P.~Herzog, K.~W.~Huang and K.~Jensen,
  %``Universal Entanglement and Boundary Geometry in Conformal Field Theory,''
  JHEP {\bf 1601}, 162 (2016)
  [arXiv:1510.00021 [hep-th]].
  %%CITATION = doi:10.1007/JHEP01(2016)162;%%
  %7 citations counted in INSPIRE as of 22 Dec 2016

 \bibitem{Herzog:2017kkj} 
  C.~Herzog, K.~W.~Huang and K.~Jensen,
  ``Displacement Operators and Constraints on Boundary Central Charges,''
  Phys.\ Rev.\ Lett.\  {\bf 120}, no. 2, 021601 (2018)
  %doi:10.1103/PhysRevLett.120.021601
  [arXiv:1709.07431 [hep-th]].
  %%CITATION = doi:10.1103/PhysRevLett.120.021601;%%
  %5 citations counted in INSPIRE as of 16 Mar 2018
  
  %\cite{Herzog:2017xha}
\bibitem{Herzog:2017xha} 
  C.~P.~Herzog and K.~W.~Huang,
  %``Boundary Conformal Field Theory and a Boundary Central Charge,''
  JHEP {\bf 1710}, 189 (2017)
  doi:10.1007/JHEP10(2017)189
  [arXiv:1707.06224 [hep-th]].
  %%CITATION = doi:10.1007/JHEP10(2017)189;%%
  %30 citations counted in INSPIRE as of 15 Jan 2019



%\cite{Jensen:2017eof}
\bibitem{Jensen:2017eof} 
  K.~Jensen, E.~Shaverin and A.~Yarom,
  ``’t Hooft anomalies and boundaries,''
  JHEP {\bf 1801}, 085 (2018)
 % doi:10.1007/JHEP01(2018)085
  [arXiv:1710.07299 [hep-th]].
  %%CITATION = doi:10.1007/JHEP01(2018)085;%%
  %1 citations counted in INSPIRE as of 16 Mar 2018

%\cite{Kurkov:2017cdz} 
\bibitem{Kurkov:2017cdz} 
  M.~Kurkov and D.~Vassilevich,
%  ``Parity anomaly in four dimensions,''
  Phys.\ Rev.\ D {\bf 96}, no. 2, 025011 (2017)
 % doi:10.1103/PhysRevD.96.025011
  [arXiv:1704.06736 [hep-th]].
  %%CITATION = doi:10.1103/PhysRevD.96.025011;%%
  %5 citations counted in INSPIRE as of 16 Mar 2018


  %\cite{Kurkov:2018pjw}
\bibitem{Kurkov:2018pjw}
M.~Kurkov and D.~Vassilevich,
%``Gravitational parity anomaly with and without boundaries,''
JHEP \textbf{03}, 072 (2018)
%doi:10.1007/JHEP03(2018)072
[arXiv:1801.02049 [hep-th]].


  %\cite{Vassilevich:2018aqu}
\bibitem{Vassilevich:2018aqu}
D.~Vassilevich,
%``Index Theorems and Domain Walls,''
JHEP \textbf{07}, 108 (2018)
%doi:10.1007/JHEP07(2018)108
[arXiv:1805.09974 [hep-th]].

%\cite{Fialkovsky:2019rum,Vassilevich:2019mhl}
\bibitem{Fialkovsky:2019rum}
I.~Fialkovsky, M.~Kurkov and D.~Vassilevich,
%``Quantum Dirac fermions in a half-space and their interaction with an electromagnetic field,''
Phys.\ Rev.\ D \textbf{100}, no.4, 045026 (2019)
%doi:10.1103/PhysRevD.100.045026
[arXiv:1906.06704 [hep-th]].

%\cite{Vassilevich:2019mhl}
\bibitem{Vassilevich:2019mhl}
D.~Vassilevich,
%``On the (im)possibility of Casimir repulsion between Chern-Simons surfaces,''
Mod.\ Phys.\ Lett.\ A \textbf{35}, no.03, 2040017 (2020)
doi:10.1142/S0217732320400179
[arXiv:1909.09049 [hep-th]].


%\cite{Rodriguez-Gomez:2017kxf} 
\bibitem{Rodriguez-Gomez:2017kxf} 
  D.~Rodriguez-Gomez and J.~G.~Russo,
%  ``Free energy and boundary anomalies on $\mathbb{S}^a\times \mathbb{H}^b$ spaces,''
  JHEP {\bf 1710}, 084 (2017)
%  doi:10.1007/JHEP10(2017)084
  [arXiv:1708.00305 [hep-th]].
  %%CITATION = doi:10.1007/JHEP10(2017)084;%%
  %3 citations counted in INSPIRE as of 16 Mar 2018
  
  %\cite{Berthiere:2019lks,FarajiAstaneh:2017hqv}
\bibitem{Berthiere:2019lks}
C.~Berthiere and W.~Witczak-Krempa,
%``Relating bulk to boundary entanglement,''
Phys. Rev. B \textbf{100}, no.23, 235112 (2019)
doi:10.1103/PhysRevB.100.235112
[arXiv:1907.11249 [cond-mat.str-el]].

%\cite{FarajiAstaneh:2017hqv}
\bibitem{FarajiAstaneh:2017hqv}
A.~Faraji Astaneh, C.~Berthiere, D.~Fursaev and S.~N.~Solodukhin,
%``Holographic calculation of entanglement entropy in the presence of boundaries,''
Phys. Rev. D \textbf{95}, no.10, 106013 (2017)
doi:10.1103/PhysRevD.95.106013
[arXiv:1703.04186 [hep-th]].
  
  %\cite{Seminara:2017hhh,Erdmenger:2015spo,Erdmenger:2014xya,Flory:2017ftd}
\bibitem{Seminara:2017hhh} 
  D.~Seminara, J.~Sisti and E.~Tonni,
 % ``Corner contributions to holographic entanglement entropy in AdS$_{4}$/BCFT$_{3}$,''
  JHEP {\bf 1711}, 076 (2017)
 % doi:10.1007/JHEP11(2017)076
  [arXiv:1708.05080 [hep-th]].
  %%CITATION = doi:10.1007/JHEP11(2017)076;%%
  %1 citations counted in INSPIRE as of 16 Mar 2018


%\cite{Erdmenger:2015spo}
\bibitem{Erdmenger:2015spo} 
  J.~Erdmenger, M.~Flory, C.~Hoyos, M.~N.~Newrzella and J.~M.~S.~Wu,
  %``Entanglement Entropy in a Holographic Kondo Model,''
  Fortsch.\ Phys.\  {\bf 64}, 109 (2016)
%  doi:10.1002/prop.201500099
  [arXiv:1511.03666 [hep-th]].
  %%CITATION = doi:10.1002/prop.201500099;%%
  %12 citations counted in INSPIRE as of 24 Jan 2017

%\cite{Erdmenger:2014xya}
\bibitem{Erdmenger:2014xya} 
  J.~Erdmenger, M.~Flory and M.~N.~Newrzella,
  %``Bending branes for DCFT in two dimensions,''
  JHEP {\bf 1501}, 058 (2015)
 % doi:10.1007/JHEP01(2015)058
  [arXiv:1410.7811 [hep-th]].
  %%CITATION = doi:10.1007/JHEP01(2015)058;%%
  %10 citations counted in INSPIRE as of 24 Jan 2017

%\cite{Flory:2017ftd}
\bibitem{Flory:2017ftd} 
  M.~Flory,
  %``A complexity/fidelity susceptibility $g$-theorem for AdS$_{3}$/BCFT$_{2}$,''
  JHEP {\bf 1706}, 131 (2017)
 % doi:10.1007/JHEP06(2017)131
  [arXiv:1702.06386 [hep-th]].
  %%CITATION = doi:10.1007/JHEP06(2017)131;%%
  %12 citations counted in INSPIRE as of 01 Jul 2018



%\cite{Miao:2017gyt,Chu:2017aab}
\bibitem{Miao:2017gyt} 
  R.~X.~Miao, C.~S.~Chu and W.~Z.~Guo,
  %``New proposal for a holographic boundary conformal field theory,''
  Phys.\ Rev.\ D {\bf 96}, no. 4, 046005 (2017)
  %doi:10.1103/PhysRevD.96.046005
  [arXiv:1701.04275 [hep-th]].
  %%CITATION = doi:10.1103/PhysRevD.96.046005;%%
  %18 citations counted in INSPIRE as of 16 Jan 2019
  
  %\cite{Chu:2017aab}
\bibitem{Chu:2017aab} 
  C.~S.~Chu, R.~X.~Miao and W.~Z.~Guo,
  %``On New Proposal for Holographic BCFT,''
  JHEP {\bf 1704}, 089 (2017)
  %doi:10.1007/JHEP04(2017)089
  [arXiv:1701.07202 [hep-th]].
  %%CITATION = doi:10.1007/JHEP04(2017)089;%%
  %14 citations counted in INSPIRE as of 16 Jan 2019
  
  %\cite{Miao:2018qkc}
\bibitem{Miao:2018qkc}
R.~Miao,
%``Holographic BCFT with Dirichlet Boundary Condition,''
JHEP \textbf{02}, 025 (2019)
%doi:10.1007/JHEP02(2019)025
[arXiv:1806.10777 [hep-th]].

%\cite{Herzog:2019bom,Herzog:2019rke}
\bibitem{Herzog:2019bom}
C.~P.~Herzog and I.~Shamir,
%``On Marginal Operators in Boundary Conformal Field Theory,''
JHEP \textbf{10}, 088 (2019)
%doi:10.1007/JHEP10(2019)088
[arXiv:1906.11281 [hep-th]].

%\cite{Herzog:2019rke}
\bibitem{Herzog:2019rke}
C.~P.~Herzog and I.~Shamir,
%``How a-type anomalies can depend on marginal couplings,''
Phys.\ Rev.\ Lett.\  \textbf{124}, no.1, 011601 (2020)
%doi:10.1103/PhysRevLett.124.011601
[arXiv:1907.04952 [hep-th]].
  
 %\cite{Jensen:2015swa,Fursaev:2015wpa,Herzog:2015ioa,Herzog:2017kkj,Herzog:2017xha,Jensen:2017eof,Kurkov:2017cdz,Kurkov:2018pjw,Vassilevich:2018aqu,Rodriguez-Gomez:2017kxf,Miao:2017gyt,Chu:2017aab}


%\cite{McAvity:1990we}
\bibitem{McAvity:1990we} 
  D.~M.~McAvity and H.~Osborn,
  %``A DeWitt expansion of the heat kernel for manifolds with a boundary,''
  Class.\ Quant.\ Grav.\  {\bf 8}, 603 (1991).
 %doi:10.1088/0264-9381/8/4/008
  %%CITATION = doi:10.1088/0264-9381/8/4/008;%%
  %64 citations counted in INSPIRE as of 22 Nov 2017
  
  %\cite{Deutsch:1978sc}
\bibitem{Deutsch:1978sc}
D.~Deutsch and P.~Candelas,
%``Boundary Effects in Quantum Field Theory,''
Phys.\ Rev.\ D \textbf{20}, 3063 (1979).
%doi:10.1103/PhysRevD.20.3063
%343 citations counted in INSPIRE as of 01 Apr 2020
%Copy to ClipboardDownload

 \bibitem{JohnP}
  John P and  Suttorp L G,
  %Boundary effects in a magnetized free-electron gas: Green function approach[J]. 
  Journal of Physics A General Physics, 1995, 28(21):6087-6097. 
  
    
%\cite{Parker:2009uva}
\bibitem{Parker:2009uva}
L.~E.~Parker and D.~Toms,
``Quantum Field Theory in Curved Spacetime,''
doi:10.1017/CBO9780511813924
  
     \bibitem{FalKovskii}
Fal'Kovskii L A,
% ``Density Attenuation of Surface Magnetic States ,''
 JETP 31, 981 (1970).  
%Fal’kovskii LA 1970 Sov. Phys.-JETP {\bf 31} 981

 

  
\bibitem{Goldhaber}
  M. Rho, A. S. Goldhaber and G. E. Brown, Phys. Rev. Lett. 51, 747 (1983).
  




  
  %\cite{Peskin:1995ev}
\bibitem{Peskin:1995ev}
M.~E.~Peskin and D.~V.~Schroeder,
``An Introduction to quantum field theory.''



\end{thebibliography}
\end{document}